# THE NANOGRAV 11-YEAR DATA SET: ARECIBO OBSERVATORY POLARIMETRY AND PULSE MICROCOMPONENTS

Peter A. Gentile,[1, 2] Maura A. McLaughlin,[1, 2] Paul B. Demorest,[3] Ingrid H. Stairs,[4] Zaven Arzoumanian,[5] Kathryn Crowter,[4] Timothy Dolch,[6] Megan E. DeCesar,[7] Justin A. Ellis,[8, 9] Robert D. Ferdman,[10] Elizabeth C. Ferrara,[11] Emmanuel Fonseca,[4] Marjorie E. Gonzalez,[4, 12] Glenn Jones,[13] Megan L. Jones,[1, 2] Michael T. Lam,[1, 2] Lina Levin,[14] Duncan R. Lorimer,[1, 2] Ryan S. Lynch,[15] Cherry Ng,[4] David J. Nice,[7] Timothy T. Pennucci,[1, 16] Scott M. Ransom,[17] Paul S. Ray,[18] Renée Spiewak,[19, 20] Kevin Stovall,[3] Joseph K. Swiggum,[19] and Weiwei Zhu[21, 22]

[1]Department of Physics and Astronomy, West Virginia University, P.O. Box 6315, Morgantown, WV 26505
[2]Center for Gravitational Waves and Cosmology, West Virginia University, Chestnut Ridge Research Building, Morgantown, WV 26505
[3]National Radio Astronomy Observatory, P.O. Box 0, Socorro, NM, 87801, USA
[4]Department of Physics and Astronomy, University of British Columbia, 6224 Agricultural Road, Vancouver, BC V6T 1Z1, Canada
[5]Center for Research and Exploration in Space Science and Technology, X-ray Astrophysics Laboratory, NASA Goddard Space Flight Center, Code 662, Greenbelt, MD 20771, USA
[6]Department of Physics, Hillsdale College, 33 E. College Street, Hillsdale, Michigan 49242, USA
[7]Department of Physics, Lafayette College, Easton, PA 18042, USA
[8]Jet Propulsion Laboratory, California Institute of Technology, 4800 Oak Grove Dr. Pasadena CA, 91109, USA
[9]Einstein Fellow
[10]Department of Physics, McGill University, 3600 rue Universite, Montreal, QC H3A 2T8, Canada
[11]NASA Goddard Space Flight Center, Greenbelt, MD 20771, USA
[12]Department of Nuclear Medicine, Vancouver Coastal Health Authority, Vancouver, BC V5Z 1M9, Canada
[13]Department of Astronomy, Columbia University, 550 W. 120th St. New York, NY 10027, USA
[14]Jodrell Bank Centre for Astrophysics, School of Physics and Astronomy, The University of Manchester, Manchester M13 9PL, UK
[15]Green Bank Observatory, P.O. Box 2, Green Bank, WV 24944, USA
[16]Institute of Physics, Eötvös Loránd University, Pázmány P. s. 1/A, Budapest 1117, Hungary
[17]National Radio Astronomy Observatory, 520 Edgemont Road, Charlottesville, VA 22903, USA
[18]Space Science Division, Naval Research Laboratory, Washington, DC 20375-5352, USA
[19]Department of Physics, University of Wisconsin-Milwaukee, Milwaukee WI 53211, USA
[20]Centre for Astrophysics and Supercomputing, Swinburne University of Technology, P.O. Box 218, Hawthorn, Victoria 3122, Australia
[21]National Astronomical Observatories, Chinese Academy of Science, 20A Datun Road, Chaoyang District, Beijing 100012, China
[22]Max-Planck-Institut für Radioastronomie, Auf dem Hügel 69, D53121, Bonn, Germany

## ABSTRACT

We present the polarization pulse profiles for 28 pulsars observed with the Arecibo Observatory by the North American Nanohertz Observatory for Gravitational Waves (NANOGrav) timing project at 2.1 GHz, 1.4 GHz, and 430 MHz. These profiles represent some of the most sensitive polarimetric millisecond pulsar profiles to date, revealing the existence of microcomponents (that is, pulse components with peak intensities much lower than the total pulse peak intensity). Although microcomponents have been detected in some pulsars previously, we present microcomponents for PSRs B1937+21, J1713+0747, and J2234+0944 for the first time. These microcomponents can have an impact on pulsar timing, geometry, and flux density determination. We present rotation measures for all 28 pulsars, determined independently at different observation frequencies and epochs, and find the Galactic magnetic fields derived from these rotation measures to be consistent with current models. These polarization profiles were made using measurement equation template matching, which allows us to generate the polarimetric response of the Arecibo Observatory on an epoch-by-epoch basis. We use this method to describe its time variability, and find that the polarimetric responses of the Arecibo Observatory's 1.4 and 2.1 GHz receivers vary significantly with time.

*Keywords:* pulsars: general, techniques: polarimetric

## 1. INTRODUCTION

Pulsars have proved themselves not only to be extremely interesting objects in and of themselves (measuring their masses can, for example, put constraints on the nuclear matter equation of state (Demorest et al. 2010)), but also incredibly useful tools that can be used to study their own local environment (see, for example Chernyakova et al. (2009)), as well as properties of the Galaxy itself (for example, pulsar observations can be used to model the interstellar medium (ISM) both broadly, as in Cordes & Lazio (2002), and coarsely, as in Coles et al. (2015) and Jones et al. (2016)). Much of the utility of pulsars stems from the ability to fit the pulse times of arrival to a timing model, while further applications are drawn from the fact that pulsars are, in general, highly polarized astrophysical signals.

The study of this polarized emission is typically described by the four Stokes parameters, $I$, $Q$, $U$, and $V$. In this vocabulary, Stokes $I$ is the total intensity of the emission, Stokes $Q$ and $U$ form the linearly polarized intensity $L$ through the relation

$$L = \sqrt{Q^2 + U^2}, \tag{1}$$

and Stokes $V$ is the circularly polarized intensity, where, using the International Astronomical Union (IAU) circular polarization sign convention, right-handed circular polarization (RCP) is positive and describes light whose position angle rotates in the counter-clockwise direction as seen by the observer. The total polarization of the emission is then

$$P = \sqrt{Q^2 + U^2 + V^2} = \sqrt{L^2 + V^2}, \tag{2}$$

Another quantity of interest is the position angle (PA) of the linearly polarized emission, which describes the orientation of the linear polarization coming from the pulsar and is given by

$$\Psi = 0.5 \tan^{-1} \frac{U}{Q}. \tag{3}$$

Note that in this convention, positive position angles are measured counter-clockwise from North. This also provides an equivalent definition of RCP: polarized light whose position angle increases with time.

The Stokes parameters describing the radiation incident on the telescope are often represented by the vector

$$S = \begin{bmatrix} I \\ Q \\ U \\ V \end{bmatrix} \tag{4}$$

which allows the effects of an imperfect receiver to be described by a 4×4 matrix as follows:

$$S_{meas} = M S_{src}, \tag{5}$$

where $S_{src}$ represents the light as emitted from the source, $S_{meas}$ represents the light as detected by the telescope, and $M$, called the Mueller Matrix, takes the form (following Lorimer & Kramer (2005). Also see Heiles (2002)):

$$M = \begin{bmatrix} 1 & E & A + EC & B + ED \\ E & H & A + EC & D + EB \\ AF - GB & GD - FC & F & -G \\ AG + BF & -GC - FD & G & FH \end{bmatrix} \tag{6}$$

where

$$
\begin{aligned}
A &= \epsilon_1 \cos(\theta_1) + \epsilon_2 \cos(\theta_2) \\
B &= \epsilon_1 \sin(\theta_1) + \epsilon_2 \sin(\theta_2) \\
C &= \epsilon_1 \cos(\theta_1) - \epsilon_2 \cos(\theta_2) \\
D &= \epsilon_1 \sin(\theta_1) - \epsilon_2 \sin(\theta_2) \\
E &= \frac{\gamma}{2} \\
F &= \cos(\phi) \\
G &= \sin(\phi) \\
H &= 1
\end{aligned}
$$

and $\epsilon_1$ and $\epsilon_2$ represent the magnitude of the cross-coupling of the two respective feeds, $\theta_1$ and $\theta_2$ represent the phase of this cross-coupling, $\gamma$ represents the differential gain of the receiver, and $\phi$ represents the differential phase of the receiver.

Observations of the Stokes parameters emitted by pulsars show that they are highly polarized sources, and that the PA of the polarized light emitted by pulsars varies with pulse phase. For many pulsars, this variation is well-described by the rotating vector model (RVM, Radhakrishnan & Cooke (1969)), which gives the pulsar's PA at a given pulse phase pulse phase (see Everett & Weisberg (2001)):

$$-\tan(\Psi - \Psi_0) = \frac{\sin\alpha \, \sin(\Phi - \Phi_0)}{\sin\zeta \, \cos\alpha + \cos\zeta \, \sin\alpha \, \cos(\Phi - \Phi_0)}, \tag{7}$$

where $\alpha$ is the angle between the rotation and magnetic axes, $\zeta$ is the angle between the rotation axis and the line of sight, and $\Psi_0$ and $\Phi_0$ are the PA and pulse phase of the inflection point of the PA swing, respectively[1]. This model assumes the pulsar radio emission is due to curvature radiation originating from charged particles flowing along dipolar magnetic field lines coming from the polar cap of the neutron star (Radhakrishnan 1969; Komesaroff 1970), and that this emission should therefore be polarized parallel to the magnetic field line along which those charged particles are streaming.

This model makes it possible to use the PA swing to characterize the geometry of the pulsar, however some pulsars, specifically millisecond pulsars, have shown vast complexity in their PA swings. These characteristics, such as orthogonal jumps in the PA swing as well as PA swings that are simply not described well by the RVM, show that while the RVM can be very useful, pulsar emission is often too complex to be described well by it.

Analyzing the variation of the Stokes parameters over quantities such as pulsar pulse phase and observing frequency can be an incredibly powerful tool for probing phenomena that are both intrinsic and extrinsic to the pulsar. For example, emission from the pulsar will be affected in various ways by the environment it travels through on its way from the pulsar to the observer. One of these ways is by Faraday rotation, which rotates the linear polarization by an angle $\beta$ determined by:

$$\beta = \frac{e^3 \lambda^2}{2 \pi m_e^2 c^4} \int_0^d n_e \, B_\parallel \, dl = \mathrm{RM} \, \lambda^2, \qquad (8)$$

where $e$ is the charge of an electron and $m_e$ is its mass, $c$ is the speed of light in a vacuum, $n_e$ is the electron number density, and $B_\parallel$ is the component of the magnetic field parallel to the direction of propagation. This can be used to measure the average $B_\parallel$ along the line of sight by computing the ratio of rotation measure RM to dispersion measure DM:

$$\langle B_\parallel \rangle = \frac{\int_0^d n_e \, B_\parallel \, dl}{\int_0^d n_e \, dl} = 1.23 \, \frac{\mathrm{RM}}{\mathrm{DM}} \, \mu\mathrm{G} \qquad (9)$$

where RM is measured in rad $\mathrm{m}^{-2}$ and DM is measured in pc $\mathrm{cm}^{-3}$.

Polarization properties of MSPs are similar in many ways to those of young pulsars. For example, they both have high polarization fractions (see Manchester et al. 1973; Dai et al. 2015; Johnston & Kerr 2018) typically dominated by linear polarization rather than circular polarization, they both display orthogonal (Backer et al. 1976; Stairs et al. 1999) and non-orthogonal (Backer & Rankin 1980; Navarro et al. 1997) jumps in the PA curve, and they both often show a reversal of the circular polarization handedness at the center of the pulse (Rankin 1983; Xilouris et al. 1998).

Despite these similarities, there are significant differences in the emission characteristics of MSPs and young pulsars. For example, the smaller rotational periods of MSPs mean that MSPs have a much smaller light cylinder radius. As a consequence, MSPs have higher duty cycles, and have profiles that are more affected by magnetic sweep-back (Barnard 1986; Kramer et al. 1999a) and relativistic aberration and retardation (Blaskiewicz et al. 1991; Dyks 2008). Further the PA swings of MSPs are typically more complex than those of young pulsars, suggesting that MSP magnetospheres are significantly non-dipolar.

Much remains to be learned from the polarization of pulsars in general and MSPs in particular. Studies of the pulsar emission mechanism (see Melrose 2004, for a review), the relationship between orthogonal modes of polarization, and the structure of the MSP magnetosphere all benefit from high signal-to-noise ratio SNR, multi-frequency polarimetric observations. In this paper, we present the polarization profiles of the pulsars observed at the Arecibo Observatory (AO) as part of NANOGrav's pulsar timing campaign. We also present the implementation of a calibration technique described in van Straten (2013), whereby a polarimetric response (PR) is created quickly, allowing PRs to be made on an epoch-by-epoch basis. In section 3 and 3.1, we describe this technique.

## 2. OBSERVATIONS

The NANOGrav Collaboration et al. (2015) describes the full details of NANOGrav's data collection process, and we summarize the relevant parts of that process here. Note that we only used data collected at the Arecibo Observatory with the wideband PUPPI[2] instrument.

Sources were observed spanning MJDs 56989 to 56874, with a quasi-monthly cadence with AO's S-Wide (center frequency 2.1 GHz), L-Wide (center frequency 1.4 GHz) and 430 MHz receivers (or some combination of the three that was chosen to most efficiently capitalize on each individual pulsar's multi-frequency characteristics). The typical integration time of a single observation was approximately 20 minutes. At 2.1 and 1.4 GHz (both dual linear feeds), the available bandwidth is 800 MHz, which is split into 512 frequency channels, while at 430 MHz (a dual circular feed), the available bandwidth is 100 MHz and is split into 64 channels.

At all observing frequencies, the data were folded and coherently dedispersed in real time. Then, frequency channels affected by RFI were excised and flux calibrator observations were used to convert telescope intensity into flux. Reported fluxes were determined by averaging fluxes over all epochs.

---

[1] The negative sign on the left side of Equation 7 represents the fact that we quote PAs using the IAU convention (positive PAs describe angles counter-clockwise from North), whereas other analyses may quote PAs in the the opposite convention.

[2] Puerto Rican Ultimate Pulsar Processing Instrument

## 3. DATA REDUCTION

### 3.1. *Polarimetric Calibration*

The determination of the parameters that make up the Mueller matrix (shown in Equation 6) is typically performed in multiple steps, the first of which is the determination of the differential gain and differential phase of the receiver ($\gamma$ and $\phi$ respectively). These parameters can be determined by injecting a correlated (that is, in phase), pulsed signal into each receiver feed and measuring the difference in gain and the phase offset induced by the receiver. Since we expect $\gamma$ and $\phi$ to vary with observing frequency, this injected signal is broad band, enabling the determination of $\gamma$ and $\phi$ for each observed frequency channel. Further, since we expect $\gamma$ and $\phi$ to also vary with time, we observed the injected signal for $\sim$90 seconds before each observation. This procedure is part of the standard NANOGrav data reduction process.

To determine the cross-coupling parameters ($\epsilon_{1,2}$ and $\theta_{1,2}$), a source with strong linear polarization is observed over a wide range of parallactic angles on the sky, which rotates the angle of polarization relative to the receiver, enabling the receiver to sample linearly polarized light at many different orientations. Equation 5 can then be expanded into four simultaneous equations which can be used to for for the remaining parameters. Since $\epsilon_{1,2}$ and $\theta_{1,2}$ are also expected to change with observing frequency, the observed calibration source should be a broad band source. There is no general assumption regarding the time variability of these sources, as major timing campaigns make assumptions ranging from assuming the cross-coupling parameters are zero (tacitly assuming they do not change over time, see Desvignes et al. (2016)), to solving for the cross-coupling parameters several times per year (Manchester et al. 2013).

To estimate the magnitude of the cross-coupling between the receiver feeds, observations of PSR J0528+2200 (PSR B0525+21), a bright, broad band, strongly polarized source, were taken over a wide range of parallactic angles. To maximize the efficiency of the observing time, the observing frequency was switched between 430 MHz, 1.4 GHz, and 2.1 GHz. This process yielded polarimetric data for each of the observing frequencies for which we have corresponding NANOGrav pulsar data, allowing us to extract three full PRs (as described in van Straten (2004)) with one observation.

These responses were then used to correct for the cross-coupling of the receiver feeds. After using these responses to calibrate our data, we then compared the resulting profiles from each of the quasi-monthly observations of PSR B1937+21 and PSR J1713+0474 to previously published profiles. These sources were chosen for comparison because they had a high SNR and many consistent published polarization profiles.

Since the polarization profiles generated by this method were not consistent epoch-to-epoch, we used PSRs J1713+0747 and B1937+21 as "standard sources" in the Measurement Equation Template Matching (METM) method, described in sections 3.1 and 5 of van Straten (2013) by choosing profiles for these sources that were consistent with previously published profiles. This allowed us to use subsequent observations of these sources to generate a new PR. In reality, since the observations being used to generate the METM PRs had already been calibrated with the MEM-generated PR, the METM-generated PRs should be thought of as per-epoch corrections to the MEM-generated PR.

Next, having already calibrated all of the observations in NANOGrav's AO data set with the MEM-generated PR, we were able to apply these corrections to these observations by selecting the PR correction whose epoch is closest to the epoch of the observation that is to be calibrated.

### 3.2. *Faraday Rotation Correction*

As described above, one of the consequences of the propagation of polarized light through the interstellar medium is the rotation of the angle of the linearly polarized emission, known as Faraday rotation. To correct for this effect, we used a range of trial RMs, where for each trial RM, the linear polarization in each profile bin of each frequency channel was rotated by the angle determined by the center frequency of the channel, the trial RM, and Equation 8. We then summed the linear polarization in frequency and pulse phase and determined the RM at which the total observed linear polarization is maximized (see Han et al. (2006)).

We found that we could reliably fit an RM value to data from most 1.4 GHz observations and some 2.1 GHz observations. For data in which RM could not be fit (430 MHz observations and some 2.1 GHz observations, because of either a low SNR, a lack of channels in which the source was strongly detected, or both), we used the RM from 1.4 GHz to correct for Faraday rotation. Errors quoted on RMs are the standard epoch to epoch variation of RM.

## 4. RESULTS

Here, we present the profiles[3] resulting from the calibration scheme described in Section 3. For each source at each observing frequency, we describe the time and frequency-averaged polarization profile and compare it to both our own results at other frequencies and to previously published results, if any exist. Profiles are shown in Figures 1-17. We note that epoch to epoch DM variations will introduce some intra-channel dispersive smearing. For all sources, the timescale for this smearing is less than the width of a profile bin given the level of variation shown in Jones et al. (2017), and is therefore negligible. The parameters describing the observations themselves (such as observation times and frequencies) and derived RMs are summarized in Table 1. Parameters describing the total intensity emission from each of the sources in this data set are shown in Table 2, where the spectral index was found by averaging the total flux over all epochs for each frequency band and "duty cycle" was defined as the ratio of bins where the total intensity emission was above the baseline (as determined by eye) to the total number of bins, rather than the definition found in Lorimer & Kramer (2005)[4]. A more detailed presentation and analysis of flux densities will be covered in another paper. Polarization emission fractions for each source in this data set are shown in Table 3 where angle brackets denote a phase-averaged quantity, and bars denote an absolute value.

---

[3] These data are available to be downloaded from data.nanograv.org/polarization.

Table 1: Total Observation Lengths and Derived Rotation Measures[a]

| PSR | Observation Time | | | 2.1 GHz RM | 1.4 GHz RM | RM-Derived B |
|---|---|---|---|---|---|---|
| | 2.1 GHz | 1.4 GHz | 430 MHz | | | |
| | | hours | | rad m$^{-2}$ | | $\mu$G |
| J0023+0923 | 0.0 | 2.5 | 3.2 | – | $-4 \pm 3$ | $-0.3 \pm 0.2$ |
| J0030+0451 | 0.0 | 6.2 | 1.5 | – | $0.5 \pm 2$ | $0.2 \pm 0.6$ |
| J1022+1001 | 0.0 | 0.0 | 0.2 | – | – | – |
| J1453+1902 | 0.5 | 0.8 | 0.8 | $13.38^\dagger$ | $13.4 \pm 0.9$ | $1.2 \pm 0.1$ |
| J1640+2224 | 0.0 | 5.2 | 3.0 | | $29 \pm 8$ | $2 \pm 0.5$ |
| J1709+2313 | 0.0 | 0.6 | 0.0 | – | $44.20^\ddagger$ | 2.15 |
| J1713+0747 | 6.5 | 8.2 | 0.0 | $15 \pm 1$ | $13 \pm 2$ | $1 \pm 0.1$ |
| J1738+0333 | 6.8 | 5.0 | 0.0 | $36^\dagger$ | $36 \pm 9$ | $1.3 \pm 0.3$ |
| J1741+1351 | 0.5 | 5.5 | 3.5 | $63^\dagger$ | $63 \pm 4$ | $3.2 \pm 0.2$ |
| J1853+1303 | 0.0 | 3.9 | 2.2 | – | $82 \pm 7$ | $3.3 \pm 0.3$ |
| B1855+09 | 0.2 | 7.2 | 6.1 | $37^\ddagger$ | $20 \pm 4$ | $1.8 \pm 0.4$ |
| J1903+0327 | 6.5 | 5.0 | 0.0 | $120 \pm 80$ | $-4 \pm 4$ | $-0.01 \pm 0.01$ |
| J1910+1256 | 5.5 | 4.9 | 0.0 | $46.39 \pm 15.90$ | $58 \pm 4$ | $1.9 \pm 0.1$ |
| J1911+1347 | 0.4 | 1.1 | 0.8 | $0.63^\ddagger$ | $-3.4 \pm 0.6$ | $-0.14 \pm 0.03$ |
| J1923+2515 | 0.0 | 3.7 | 1.9 | – | $7 \pm 9$ | $0.5 \pm 0.6$ |
| B1937+21 | 4.6 | 3.7 | 0.0 | $9.22 \pm 1.96$ | $10 \pm 1$ | $0.17 \pm 0.02$ |
| J1944+0907 | 0.2 | 5.6 | 3.4 | $-30^\ddagger$ | $-30 \pm 10$ | $-1.7 \pm 0.5$ |
| J1949+3106 | 4.1 | 6.3 | 0.0 | $212^\dagger$ | $212 \pm 7$ | $1.59 \pm 0.05$ |
| B1953+29 | 0.0 | 5.0 | 2.6 | – | $8 \pm 2$ | $0.10 \pm 0.03$ |
| J1955+2527 | 0.0 | 1.5 | 0.0 | – | $-120 \pm 60$ | $-0.7 \pm 0.3$ |
| J2017+0603 | 6.4 | 9.7 | 2.0 | $-41 \pm 8$ | $-56 \pm 5$ | $-2.9 \pm 0.3$ |
| J2019+2425 | 0.0 | 0.6 | 0.7 | – | $-71 \pm 4$ | $-5.1 \pm 0.3$ |
| J2033+1734 | 0.0 | 1.2 | 0.8 | – | $-69.4 \pm 0.3$ | $-3.41 \pm 0.02$ |
| J2043+1711 | 0.0 | 10.9 | 6.5 | – | $-69 \pm 4$ | $-4.1 \pm 0.2$ |
| J2214+3000 | 4.5 | 4.5 | 0.0 | $-60 \pm 13$ | $-43 \pm 5$ | $-2.4 \pm 0.3$ |
| J2229+2643 | 0.0 | 1.4 | 1.1 | – | $-58 \pm 2$ | $-3.1 \pm 0.1$ |
| J2234+0611 | 0.0 | 2.0 | 0.0 | – | $-2 \pm 1$ | $-0.2 \pm 0.1$ |
| J2234+0944 | 0.8 | 1.3 | 1.0 | $-6 \pm 42$ | $-8 \pm 4$ | $-0.5 \pm 0.3$ |
| J2317+1439 | 0.0 | 6.8 | 7.7 | – | $-9 \pm 4$ | $-0.5 \pm 0.2$ |

[a]Errors on RM represent the standard epoch to epoch deviation of RM. Values with a $\dagger$ have no error because the data were too low SNR, so the average 1.4 GHz RM was used. Values with a $\ddagger$ have no error because there was only one observation at this frequency.

Table 2: Total Intensity Emission Parameters

| PSR | Flux Density | | | Duty Cycle | | | Spectral Index |
|-----|---------|---------|---------|---------|---------|---------|---------|
| | 2.1 GHz | 1.4 GHz | 430 MHz | 2.1 GHz | 1.4 GHz | 430 MHz | |
| | mJy | mJy | mJy | | | | |
| J0023+0923 | – | 0.43 | 5.69 | – | 0.30 | 0.30 | −2.19 |
| J0030+0451 | – | 1.19 | 14.85 | – | 0.54 | 0.60 | −2.14 |
| J1022+1001 | – | – | 10.95 | – | – | 0.17 | – |
| J1453+1902 | 0.04 | 0.09 | 1.58 | 0.15 | 0.29 | 0.50 | −2.19 |
| J1640+2224 | – | 0.69 | 18.71 | – | 0.25 | 0.28 | −2.80 |
| J1709+2313 | – | 0.12 | – | – | 0.34 | – | – |
| J1713+0747 | 6.15 | 11.52 | – | 0.77 | 0.89 | – | −1.16 |
| J1738+0333 | 0.59 | 0.71 | – | 0.36 | 0.37 | – | −0.34 |
| J1741+1351 | 0.03 | 0.45 | 4.10 | 0.03 | 0.38 | 0.41 | −2.70 |
| J1853+1303 | – | 0.52 | 6.91 | – | 0.55 | 0.38 | −2.19 |
| B1855+09 | 3.38 | 3.81 | 18.06 | 0.38 | 0.75 | 0.71 | −1.03 |
| J1903+0327 | 0.68 | 0.74 | – | 0.22 | 0.63 | – | −0.16 |
| J1910+1256 | 0.30 | 0.52 | – | 0.18 | 0.31 | – | −1.02 |
| J1911+1347 | 0.93 | 0.99 | 1.99 | 0.54 | 0.53 | 0.46 | −0.47 |
| J1923+2515 | – | 0.26 | 4.03 | – | 0.59 | 0.32 | −2.32 |
| B1937+21 | 7.07 | 13.19 | – | 0.82 | 0.85 | – | −1.16 |
| J1944+0907 | 0.53 | 2.80 | 32.43 | 0.48 | 0.75 | 0.74 | −2.34 |
| J1949+3106 | 0.07 | 0.13 | – | 0.07 | 0.10 | – | −1.15 |
| B1953+29 | – | 0.89 | 26.68 | – | 0.78 | 0.50 | −2.88 |
| J1955+2527 | – | 0.39 | – | – | 0.26 | – | – |
| J2017+0603 | 0.29 | 0.35 | 1.96 | 0.35 | 0.62 | 0.54 | −1.17 |
| J2019+2425 | – | 0.28 | 5.89 | – | 0.48 | 0.35 | −2.58 |
| J2033+1734 | – | 0.41 | 2.82 | – | 0.45 | 0.45 | −1.63 |
| J2043+1711 | – | 0.20 | 4.14 | – | 0.82 | 0.41 | −2.57 |
| J2214+3000 | 0.80 | 0.61 | – | 0.47 | 0.60 | – | 0.50 |
| J2229+2643 | – | 0.71 | 5.48 | – | 0.27 | 0.28 | −1.73 |
| J2234+0611 | – | 1.28 | – | – | 0.55 | – | – |
| J2234+0944 | 1.11 | 2.05 | 3.07 | 0.50 | 0.82 | 0.45 | −0.55 |
| J2317+1439 | – | 0.75 | 45.99 | – | 0.45 | 0.23 | −3.49 |

Also, we note the presence of instrumental effects in the data. For example, the linearly polarized emission for some sources exceeds the total intensity emission in some phase bins. In some instances, this is caused by the linear polarization noise being biased by virtue of it being a quantity derived from a quadrature sum, while in other cases, it is due to baselining effects. In these latter cases, the linearly polarized emission in said phase bins should be viewed as an aberration. Additionally, for certain bright sources, sampling effects cause the pulsar signal to be aliased back into the band. Therefore, some of the low-intensity emission seen in the profiles presented here is not intrinsic, but rather due to instrumental effects (see Pennucci (2015) for more details on this issue). We stress, though, that while this effect accounts for some of the low-intensity emission, we do see additional low-intensity emission that is not due to this effect and is therefore intrinsic to the source itself.

## 4.1. PSR J0023+0923

### 4.1.1. 1.4 GHz

The 1.4 GHz polarization profile for PSR J0023+0923 published in Craig (2014) shows many similarities to our results in that the total intensity profile shows a leading component that has a relatively low intensity followed by two relatively high intensity components. The linear intensity profile also shows three components as well as evidence for a bridge of linearly polarized emission connecting them all. Although the leading component again has a lower intensity than the trailing two components, the disparity is not as pronounced as in the total intensity profile. Also of note is that the final component in the linear polarization profile is noticeably nar-rower than the component immediately preceding it, whereas the widths of their total intensity counterparts are not as disparate.

For all the similarities our results show, there are some differences. First, the circular polarization profile shows a left circularly polarized (LCP) peak that coincides with the final total intensity component, as well as evidence for a right circularly polarized (RCP) peak immediately preceding it. There is also evidence for an LCP peak coinciding roughly with the middle total intensity component, and a RCP peak coinciding with the leading total intensity component. It appears as though this would be consistent with Craig (2014) if they had also shown negative circular polarization intensities, modulo an overall negative sign. This suggests that they are using the IAU convention for circular polarization, although the circular polarization sign convention used in the paper is not stated. In addition, the total intensity profile they present shows that the final component has a significantly higher intensity than the component immediately preceding it. As the relative intensity of these components changes substantially with frequency, it is likely that the pulsar was detected much more strongly in the lower part of the observing bandwidth (where the final component is strongest) during the observations taken for the analysis presented in Craig (2014) due to interstellar scintillation.

### 4.1.2. 430 MHz

No polarization profiles have been published for PSR J0023+0923 at 430 MHz, however Bangale et al. (in prep) presents the total intensity profile as seen by the 350 MHz receiver on the Green Bank Telescope, whose frequency range overlaps with the AO 430 MHz receiver. Both profiles show a very narrow pulse, however what appears to be a single component profile in the published Green Bank Telescope profile is shown to be a multi-component main pulse consisting of an extremely bright component preceded by a low-intensity peak and followed by a moderate-intensity ridge. This main pulse is preceded by a low-intensity pulse that does not appear to be connected to the main pulse and is aligned with the precursor seen at 1.4 GHz.

The profile is not strongly polarized, with the only significant features being a peak in both the linear polarization and the right-circular polarization coincident with the total intensity peak.

## 4.2. PSR J0030+0451

### 4.2.1. 1.4 GHz

PSR J0030+0451 has a complex profile at 1.4 GHz, drawing interest even upon discovery (Lommen et al. 2000). Their analysis found a total intensity profile consisting of at least six components. Our results largely agree with this analysis, however the significant increase in signal-to-noise provided by this work reveals that the interpulse may be more complex than the two component model described in Lommen et al. (2000). We also do not see the "bump" situated between the main pulse and the interpulse (corresponding to a rotational phase of $\sim 0.8$ in our analysis), although both components appear to be wider than originally reported.

---

[4] We have adpoted this definition of "duty cycle" because the high SNR of the dataset has allowed the detection of some profile components hundreds of times smaller than the main profile peak, and "duty cycle" as defined in Lorimer & Kramer (2005) is not sensitive to those profile components.

The linear polarization profile of PSR J0030+0451 is similar in shape to that of the total intensity profile, although the main pulse shows two low-intensity linear polarization peaks at the leading edge of the main pulse and another low-intensity linear polarization peak on the trailing edge of the main pulse. The interpulse shows three distinct components that coincide with three components of the total intensity profile.

The circular polarization profile shows a strong LCP peak coinciding with the total intensity peak, along with lower-intensity peaks in RCP and LCP following it, The interpulse shows almost no circular polarization, although there is evidence for a wide, RCP pulse spanning almost the entire interpulse width.

#### 4.2.2. *430 MHz*

The profile for PSR J0030+0451 at 430 MHz is similar to the profile at 1.4 GHz, where the total intensity profile can also be seen to have two main components, although the leading component appears to be wider at 1.4 GHz than at 430 MHz. We also find the polarization profile to be less complex at 430 MHz than at 1.4 GHz. While this polarization profile is in good agreement with that presented in Lommen et al. (2000), there are small differences. As previously reported, the linear polarization profile of the main pulse has many components, with the overall peak coinciding with the total intensity peak, however our data do not show the linear polarization intensity dropping off as steeply afterwards. The circular polarization profile is also largely in agreement with that shown in Lommen et al. (2000), with the only significant difference being the RCP peak presented in this work is more significant.

### 4.3. *PSR J1022+1001*

#### 4.3.1. *430 MHz*

PSR J1022+1001 profile has proven to be variable (see Kramer et al. (1999b); Kuz'min & Losovskii (1999), and Kramer et al. (1999a)), most notably with the respective intensities of the two components that make up the total intensity profile. In this work, we see the leading component have a higher intensity than the trailing component, consistent with some later observations (Stairs et al. (1999), for example), but not with its discovery Camilo et al. (1996a). The linearly polarized emission also shows two primary components, the first of which precedes its total intensity counterpart by a phase of $\sim 0.02$, while the other is aligned with its total intensity counterpart. The circularly polarized emission is entirely LCP and also shows two components, although unlike the linearly polarized emission, these components are aligned with their total intensity counterparts.

### 4.4. *PSR J1453+1902*

#### 4.4.1. *2.1 GHz*

The SNR of the 2.1 GHz profile for PSR J1453+1902 is not very high, but one component is clearly detected. That component shows moderate linear polarization, which also consists of one component, the center of which precedes the main total intensity component. We do not detect any circularly polarized emission.

#### 4.4.2. *1.4 GHz*

Much of PSR J1453+1902's total intensity profile at 1.4 GHz resembles its 2.1 GHz counterpart, as the main part of the profile consists of one bright component. However, the higher SNR at 1.4 GHz reveals an additional component leading the brightest component. As with the 2.1 GHz profile, there is a bright linearly polarized peak of emission that precedes the total intensity peak by a phase of $\sim 0.3$, although it appears as though there is another, very weak component trailing that. The new total intensity component is 100% polarized. As with the 2.1 GHz profile, we do not detect significant circular polarization.

#### 4.4.3. *430 MHz*

The only published profile for PSR J1453+1902 is a 430 MHz total intensity profile that agrees well with our 430 MHz profile (Lorimer et al. 2007), showing a leading, low intensity component, followed by much stronger component that has a faint, trailing pulse attached to it. The flux density of the total intensity profile is also in agreement with the analysis done in Lorimer et al. (2007). The linear polarization profile has changed markedly from the 2.4 and 1.4 GHz linear intensity profiles, showing one, low intensity component that appears to be aligned with the total intensity peak. The circularly polarized profile is also very different at 430 MHz, where there is a clear LCP peak aligned with the total intensity peak, whereas at higher frequencies, there is no detectable circular polarization.

### 4.5. *PSR J1640+2224*

#### 4.5.1. *1.4 GHz*

PSR J1640+2224's pulse profile at 1.4 GHz is a single pulse comprised of at least three components (Kramer et al. 1998). The linear polarization profile shows four distinct components, the brightest of which coincides with the total intensity peak. PSR J1640+2224 also shows a strong LCP peak (with a higher intensity than the linear polarization peak) coincident with the total intensity peak, and flanked by low-intensity RCP emission.

#### 4.5.2. *430 MHz*

PSR J1640+2224's pulse profile at 430 MHz is similar to the profile at 1.4 GHz, as the components on the 1.4 GHz profile are all still present and the total profile widths are very similar (Foster et al. 1995). The linear polarization profile shows two main components that coincide with two components of the total intensity profile, and a trailing linear polarization tail. The circular polarization is wholly RCP and is nearly identical to the linear polarization profile, although with a peak intensity $\sim 0.8$ times that of the peak linearly polarized intensity.

### 4.6. *PSR J1709+2313*

#### 4.6.1. *1.4 GHz*

The 1.4 GHz profile of PSR J1709+2313 presented here is in agreement with the previously published profile (Lewandowski et al. 2004), and includes a main pulse with at least 3 components in addition to an interpulse. The interpulse is almost entirely linearly polarized, as is the leading component of the main pulse. There is no detectable circular polarization.

### 4.7. *PSR J1713+0747*

#### 4.7.1. *2.1 GHz*

The pulse profile for PSR J1713+0747 at 2.1 GHz has been studied before (Dai et al. 2015), and our observations confirm the analysis presented therein. The increased sensitivity of our data set, however, allows us to detect small components preceding and trailing the main total intensity profile components. These microcomponents are 50 to 100 times less intense than the total intensity peak, respectively, yet almost double the width of the pulse. The leading component shows slight linear polarization, yet because of its relatively low total intensity, this small level of polarization means that the leading component is completely polarized. The trailing component, however, does not show any detectable polarized emission.

#### 4.7.2. *1.4 GHz*

PSR J1713+0747 has been studied extensively at 1.4 GHz, and in terms of the most intense components, we again find agreement with the existing literature (see, for example Dai et al. (2015); Ord et al. (2004); Yan et al. (2011), and more). Yet as with the 2.1 GHz data, we find substantial structure with intensities $\sim$100 times smaller than the peak total intensity. Indeed, we see the same microcomponents that were evident in the 2.1 GHz emission, although they are much higher SNR at 1.4 GHz. Also, what appeared to be a short tail preceding the first total intensity microcomponent appears here to be a much longer tail, spanning nearly the entire rotation of the pulsar.

### 4.8. *PSR J1738+0333*

#### 4.8.1. *2.1 GHz*

The 2.1 GHz total intensity profile for PSR J1738+0333 consists of one main pulse made up of two strong components preceded by two low-intensity components. The linear polarization profile shows many similarities to the total intensity profile, having 3 distinct components that correspond to components of the total intensity profile. The circular polarization profile has one strong component, and one weakly detected component. Both of these components are RCP and correspond to the strong total intensity components.

#### 4.8.2. *1.4 GHz*

The 1.4 GHz profile for PSR J1738+0333 presented in this work looks very similar not only to previously published profiles (Jacoby 2005; Freire et al. 2012), but also to its 2.1 GHz profile. The linear polarization profile has 3 distinct components that have corresponding components in the total intensity profile and the circular polarization profile consists of two components of all RCP emission that coincide to the strongest components of the total intensity profile. Indeed, the only new feature that was not seen in the 2.1 GHz data is a very weak total intensity component on the trailing edge of the main pulse.

### 4.9. *PSR J1741+1351*

#### 4.9.1. *2.1 GHz*

The 2.1 GHz profile for PSR J1741+1351 is relatively simple: one bright, single-component pulse. There does seem to be another component preceding the main pulse by 0.3 rotations, but this component is very weak, and therefore cannot be characterized further. The polarization profile for PSR J1741+1351 at 2.1 GHz contains no detectable polarized emission.

#### 4.9.2. *1.4 GHz*

The 1.4 GHz profile for PSR J1741+1351 shows the low-intensity component seen in the 2.1 GHz in much greater detail. In addition, the bright, single-component pulse seen at 2.1 GHz is not seen to consist of one bright component preceded by a low-intensity "bump" and followed by a long, low-intensity tail. This tail was not reported in previously published studies (Jacoby et al. 2007; Espinoza et al. 2013), The linearly polarized intensity profile shows that the leading part of the profile's linearly polarized emission also has two components that coincide with two components of the total intensity emission. The trailing part of the linearly polarized intensity profile has three clear components, two of which coincide with components of the total intensity profile. There is no detectable linear polarization that coincides with the low-intensity tail seen in the total intensity profile. The circular polarization profile shows a LCP pulse corresponding to the leading part of the total intensity profile, followed by a multi-component section with many changes in senses of circular polarization.

#### 4.9.3. *430 MHz*

As expected (Espinoza et al. 2013), PSR J1741+1351's 430 MHz total intensity profile profile looks similar to the 1.4 GHz profile, however, the components immediately leading and following the strongest component are different. The former is much stronger relative to the total intensity peak, whereas the latter (seen at 1.4 GHz as a long tail) is seen as two distinct components. The linear polarization profile has changed significantly. Indeed, there is almost no detectable linearly polarized emission, save a small peak of emission coinciding with the total intensity peak. The circular polarization profile shows two RCP components, one coinciding with the total intensity peak, and one coinciding with the total intensity component preceding it.

### 4.10. *PSR J1853+1303*

#### 4.10.1. *1.4 GHz*

The overall shape of PSR J1853+1303 is quite complex, and spans more than half of the rotation of the pulsar (Gonzalez et al. 2011; Stairs et al. 2005). This work shows a clear bridge of emission spanning the two major components of the profile. In light of this complexity, it is no surprise that the linear polarization profile shows significant complexity as well, containing six distinct components. Interestingly, only two of them seem to coincide with components found in the total intensity profile. Further, the two components that do have corresponding components of the total intensity profile are of relatively low intensity. The circular polarization profile, on the other hand, contains a strong LCP peak flanked by two strong RCP peaks, and they all appear to coincide with total intensity components.

#### 4.10.2. *430 MHz*

The 430 MHz profile for PSR J1853+1303 shows similar complexity to the 1.4 GHz profile, however some differences are apparent. For example, the relative brightnesses of the three brightest components have changed, and the bridge of emission connecting the two main parts of the pulse profile is no longer detectable above the noise. For as much as the total intensity profile has changed, though, the linear polarization profile has changed wholly. One broad pulse spans the first part of the profile, and there is no detectable linear intensity emission coincident with the second park of the pulse profile. The circular polarization profile has undergone a similar transformation: the strong LCP pulse seen at 1.4 GHz has turned into a strong RCP pulse, and the RCP pulse following it has turned into a weak LCP pulse. In addition, some weak circularly polarized emission is coincident with the trailing part of the profile.

### 4.11. *PSR B1855+09*

#### 4.11.1. *1.4 GHz*

The 1.4 GHz pulse profile has been well studied (Kramer et al. (1998), Yan et al. (2011), Dai et al. (2015), and many more) and shows a total intensity profile that is consistent with our work, however, we clearly detect a bridge of emission connecting the two pulses that make up the profile. This is evidence that PSR B1855+09 is an aligned rotator, a question which has generated much discussion and disagreement (see, for example, Segelstein et al. (1986), Thorsett & Stinebring (1990), and Rankin (1990)).

The linearly polarized emission presented here is in agreement with previously published profiles (see Yan et al. (2011) and Dai et al. (2015)), which shows relatively low levels of linear polarization with components that coincide with all of the components in the total intensity profile, however, we are able to resolve multiple additional components. The circularly polarized emission we present is also largely in agreement with published profiles, although we note that we detect less circularly polarized intensity.

#### 4.11.2. *430 MHz*

Just as at 1.4 GHz, the total intensity profile for PSR B1855+09 at 430 MHz (Thorsett & Stinebring 1990) also shows two main pulses. With the improved sensitivity of this data set, we are able to detect every component in the 1.4 GHz total intensity profile at 430 MHz, with the exception of the bridge of emission. The polarized emission, on the other hand, looks very different at 430 MHz than at 1.4 GHz, as the degree of linear polarization is lower in general. The linearly polarized emission does seem to share some of the same properties as the 1.4 GHz emission, as there are some distinguishable components that have corresponding total intensity components, but the circularly polarized emission looks very different. The only detectable phenomenology is weak RCP emission over much of both pulses.

### 4.12. *PSR J1903+0327*

#### 4.12.1. *2.1 GHz*

PSR J1903+0327's 2.1 GHz total intensity profile shows three components: a bright main component flanked by two weaker components. Measurements of interstellar scattering presented in Levin et al. (2016) suggest that scattering should be small at 2.1 GHz. This is consistent with PSR J1903+0327's 2.1 GHz total intensity profile, as it is difficult to identify profile components clearly affected by interstellar scattering. The linearly polarized emission shows that the weak, leading component is almost fully polarized, while the main component is coincident with the strongest linearly polarized component. There is also a moderate, RCP component coincident with the main pulse peak, which is the only detectable circularly polarized emission.

Table 3: Polarized Intensity Parameters

| PSR | ⟨P⟩/I | | | ⟨L⟩/I | | | ⟨V⟩/I | | | ⟨\|V\|⟩/I | | |
|---|---|---|---|---|---|---|---|---|---|---|---|---|
| | 2.1 GHz | 1.4 GHz | 430 MHz | 2.1 GHz | 1.4 GHz | 430 MHz | 2.1 GHz | 1.4 GHz | 430 MHz | 2.1 GHz | 1.4 GHz | 430 MHz |
| | | | | | | Percent | | | | | | |
| J0023+0923 | – | 0.31 | 0.36 | – | 0.24 | 0.03 | – | 0.03 | −0.06 | – | 0.06 | 0.18 |
| J0030+0451 | – | 0.34 | 0.25 | – | 0.31 | 0.15 | – | 0.01 | −0.05 | – | 0.04 | 0.08 |
| J1022+1001 | – | – | 0.48 | – | – | 0.26 | – | – | 0.22 | – | – | 0.26 |
| J1453+1902 | 0.67 | 0.86 | 0.82 | 0.27 | 0.37 | 0.03 | 0.07 | −0.03 | 0.24 | 0.21 | 0.27 | 0.45 |
| J1640+2224 | – | 0.16 | 0.23 | – | 0.11 | 0.16 | – | 0.02 | −0.10 | – | 0.08 | 0.11 |
| J1709+2313 | – | 0.61 | – | – | 0.13 | – | – | 0.05 | – | – | 0.25 | – |
| J1713+0747 | 0.32 | 0.33 | – | – | 0.30 | 0.32 | – | −0.02 | −0.01 | – | 0.03 | 0.03 |
| J1738+0333 | 0.30 | 0.23 | – | – | 0.10 | 0.19 | – | −0.05 | −0.02 | – | 0.13 | 0.04 |
| J1741+1351 | 0.33 | 0.22 | 0.23 | 0.07 | 0.17 | 0.01 | −0.09 | 0.01 | −0.07 | 0.12 | 0.05 | 0.13 |
| J1853+1303 | – | 0.33 | 0.35 | – | 0.18 | 0.05 | – | 0.03 | −0.09 | – | 0.19 | 0.19 |
| B1855+09 | 0.23 | 0.15 | 0.16 | 0.07 | 0.12 | 0.02 | −0.03 | −0.01 | −0.03 | 0.09 | 0.05 | 0.08 |
| J1903+0327 | 0.23 | 0.21 | – | 0.04 | 0.13 | – | −0.08 | −0.06 | – | 0.14 | 0.08 | – |
| J1910+1256 | 0.27 | 0.25 | – | 0.12 | 0.16 | – | 0.01 | −0.00 | – | 0.15 | 0.14 | – |
| J1911+1347 | 0.53 | 0.44 | 0.38 | 0.19 | 0.28 | 0.03 | 0.15 | 0.18 | 0.01 | 0.27 | 0.23 | 0.21 |
| J1923+2515 | – | 0.33 | 0.30 | – | 0.16 | 0.04 | – | −0.03 | −0.13 | – | 0.11 | 0.18 |
| B1937+21 | 0.30 | 0.31 | – | 0.28 | 0.31 | – | −0.01 | 0.00 | – | 0.04 | 0.02 | – |
| J1944+0907 | 0.60 | 0.16 | 0.21 | 0.00 | 0.10 | 0.03 | 0.01 | 0.07 | −0.13 | 0.28 | 0.08 | 0.15 |
| J1949+3106 | 0.31 | 0.17 | – | 0.07 | 0.11 | – | −0.01 | 0.01 | – | 0.13 | 0.06 | – |
| B1953+29 | – | 0.35 | 0.20 | – | 0.18 | 0.05 | – | −0.08 | −0.03 | – | 0.23 | 0.10 |
| J1955+2527 | – | 0.15 | – | – | 0.02 | – | – | −0.08 | – | – | 0.10 | – |
| J2017+0603 | 0.68 | 0.40 | 0.76 | 0.13 | 0.31 | 0.01 | −0.01 | −0.02 | −0.02 | 0.29 | 0.07 | 0.39 |
| J2019+2425 | – | 0.52 | 0.42 | – | 0.17 | 0.05 | – | 0.00 | 0.14 | – | 0.19 | 0.24 |
| J2033+1734 | – | 0.43 | 0.39 | – | 0.27 | 0.03 | – | −0.01 | −0.07 | – | 0.15 | 0.22 |
| J2043+1711 | – | 0.62 | 0.49 | – | 0.42 | 0.12 | – | 0.03 | −0.24 | – | 0.13 | 0.34 |
| J2214+3000 | 0.37 | 0.39 | – | 0.17 | 0.34 | – | −0.03 | −0.00 | – | 0.12 | 0.03 | – |
| J2229+2643 | – | 0.24 | 0.41 | – | 0.19 | 0.01 | – | 0.03 | 0.08 | – | 0.07 | 0.21 |
| J2234+0611 | – | 0.33 | – | – | 0.29 | – | – | 0.03 | – | – | 0.05 | – |
| J2234+0944 | 0.25 | 0.20 | 0.78 | 0.08 | 0.13 | 0.08 | 0.07 | 0.06 | 0.01 | 0.13 | 0.08 | 0.37 |
| J2317+1439 | – | 0.29 | 0.14 | – | 0.24 | 0.07 | – | 0.06 | −0.04 | – | 0.07 | 0.09 |

#### 4.12.2. *1.4 GHz*

PSR J1903+0327's 1.4 GHz total intensity profile is very similar to the 2.1 GHz total intensity profile, showing the same three components. The linearly polarized emission is not strikingly different either, although at 1.4 GHz, the leading component shows a very low polarization fraction. Also, the brightest linearly polarized component, while still coincident with the total intensity peak, appears to be much wider that at 2.1 GHz. The circularly polarized emission appears to be nearly identical at 1.4 GHz, showing a RCP component coincident with the total intensity peak. Unlike PSR J1903+0327's 2.1 GHz profile, the effects of scattering can be clearly identified, as not only does the leading components appear to be scattered into the main component, but the trailing components is followed by a long scattering tail.

### 4.13. *PSR J1910+1256*

#### 4.13.1. *2.1 GHz*

The 2.1 GHz profile for PSR J1910+1256 consists of one pulse with one bright, primary component preceded by a low-intensity tail and followed by a small bump leading into an-other long tail. The longer, trailing tail is almost completely polarized, while the leading trail does not show evidence for polarized emission. The remainder of the linear intensity profile consists of two components that are roughly, though not exactly, coincident with the bright, primary total intensity component. The circularly polarized profile contains one bright LCP peak that is coincident with the brightest linear polarization peak, followed by a reversal in the sense of circularly polarized emission, leading to a RCP peak which trails off into a long tail.

### 4.13.2. *1.4 GHz*

PSR J1910+1256's 1.4 GHz profile (Stairs et al. 2005; Lorimer et al. 2006; Gonzalez et al. 2011) looks very similar to its 2.1 GHz profile in that there is still a one bright, primary component preceded by a low-intensity tail and followed by a small bump leading into another long tail, and indeed both the linear and circular polarization profiles share many of the same overall characteristics of the 2.1 GHz polarized emission. For example, the linearly polarized emission still shows two main components roughly, yet not exactly coinciding with the brightest total intensity component and emission that extends into the trailing tail, and the circularly polarized emission still contains one bright LCP peak that is coincident with the brightest linear polarization peak, followed by a reversal in the sense of circularly polarized emission, leading to a RCP peak which trails of into a long tail, yet there are some small, but nonetheless interesting differences. For example, the linear intensity profile now shows a very small component preceding the two main components, and the long linearly polarized tail of the 2.1 GHz emission is seen at 1.4 GHz to be two components, the first of which roughly aligns with the bump seen in the total intensity profile. Also of note is that the ratio of peak linearly polarized intensity to peak circularly polarized intensity has reversed: at 2.1 GHz, the circularly polarized peak was more intense than the linearly polarized peak, whereas the opposite is true at 1.4 GHz.

### 4.14. *PSR J1911+1347*

#### 4.14.1. *2.1 GHz*

The 2.1 GHz total intensity profile of PSR J1911+1347 consists of a very bright pulse neighbored on either side by lower intensity components. Inspection of these lower-intensity components reveals that the final component of the total intensity profile appears to be connected to the rest of the profile. The linear polarization profile shows that this component is nearly 100% polarized, and that there is a strong linearly polarized component coincident with the total intensity peak. Also coincident with the total intensity peak is a strong LCP peak, followed by a sense reversal and a very weak RCP component. There also appears to be another, very weak RCP peak coincident with the final total intensity component.

#### 4.14.2. *1.4 GHz*

PSR J1911+1347's 1.4 GHz total intensity profile (Lorimer et al. 2006) is strikingly similar to its 2.1 GHz counterpart, and we show that bridge of emission connecting the final component to the rest of the total intensity profile seen at 2.1 GHz is even more evident at 1.4 GHz. The linear intensity profile at 1.4 GHz is nearly identical to the 2.1 GHz linear intensity profile, save the preceding components that lead the linearly polarized peak. Unsurprisingly, the circularly polarized emission has also remained largely unchanged, with the weak RCP components being even more evident.

#### 4.14.3. *430 MHz*

PSR J1911+1347's 430 MHz profile does not look wholly unlike its 1.4 GHz profile as the primary component appears to be mostly unchanged, however what was seen as a long precursor tail appears as two separate components at 430 MHz. We do not detect linear polarization, but we do detect circular polarization, which shows a LCP peak coinciding with the main pulse and is followed by a possible sense reversal and RCP peak, although these features are extremely weak.

### 4.15. *PSR J1923+2515*

#### 4.15.1. *1.4 GHz*

The 1.4 GHz profile for PSR J1923+2515 that was reported upon its discovery (Lynch et al. 2013) showed a fairly simple, two-component pulse, however the detection of an additional component preceding the 2.1 GHz total intensity peak by $\sim 0.35$ rotations coupled with the detection of that same additional component at 820 MHz suggests that increased sensitivity may show more detail in PSR J1923+2515's 1.4 GHz emission than was originally reported. The improved sensitivity afforded by AO reveals that and other lower-intensity components. In fact, there is evidence for another component preceding the aforementioned feature, as well as yet another component following the aforementioned feature (the latter of which is also seen in the 2.1 GHz profile).

The linear polarization profile shows many components that correspond to components in the total intensity profile, including three major components that roughly align with the two major components of the total intensity profile. These components are connected with a bridge of linearly polarized emission. PSR J1923+2515 does not show very much circularly polarized emission, however, there are two RCP components that align with two corresponding linearly polarized components.

#### 4.15.2. *430 MHz*

The 430 MHz profile for PSR J1923+2515 shows a similar morphology to the 1.4 GHz profile in that there are two main components preceded by a weak component (the same component seen at 2.1 and 820 MHz in Lynch et al. (2013)), meaning that this component is seen consistently from 430 MHz to 2.1 GHz. At 2.1 GHz, the first of the two main components is much brighter than the trailing component, and as the frequency of observation decreases, so does the intensity of the first main component with respect to the second. We show that this trend continues at 430 MHz.

The linear polarization profile shows one component with a long, leading tail and a peak that corresponds to the trailing component of the total intensity profile. The circular polarization profile appears to mirror the linear polarization profile, although in place of a leading tail, there are three RCP components whose intensities increase with pulse phase.

### 4.16. *PSR B1937+21*

#### 4.16.1. *2.1 GHz*

The most sensitive 2.1 GHz polarization profile for PSR B1937+21 was presented in Dai et al. (2015), however, the profiles were significantly affected by dispersive smearing, an issue which is not shared by the profiles presented here. For example, we see the component following the total intensity peak that is commonly seen in coherently dedispersed PSR B1937+21 profiles at multiple frequencies. We also see the very complex microcomponents in the total intensity profiles, except we show another, very faint component after the main pulse that does not appear in previously published profiles. We also see that the microcomponent preceding the total intensity peak by ∼0.75−0.8 is in fact two components, and are able to resolve the microcomponent centered at a pulse phase of ∼0.6 enough to see that it is in fact three components.

The linearly polarized emission at 2.1 GHz shows that the main pulse is comprised of three components which, like the corresponding total intensity pulse, is preceded by a long tail. The lower-intensity pulse has two linearly polarized components, however the second of those components is extremely weak. The microcomponents show significant polarization fractions, especially the microcomponents spanning pulse phases of ∼0.6−0.8. The microcomponents do not show any detectable circularly polarized emission, while the lower-intensity pulse has a coincident RCP peak, and the main pulse has a RCP peak, followed by a reversal in the sense of circular polarization coincident with the total intensity peak, then a LCP peak.

#### 4.16.2. *1.4 GHz*

The polarization profile of PSR B1937+21 has been studied extensively at 1.4 GHz (see, for example Dai et al. (2015); Ord et al. (2004); Yan et al. (2011), and more), and our results are broadly in good agreement with that consensus, although we are able to resolve the rich microcomponents with coherently dedispersed data for the first time. We see the same microcomponents detected at 1.4 GHz, although with much higher SNR. In fact, we see that the microcomponent occurring at a pulse phase of ∼0.8 appears to be two components. The increased SNR at 1.4 GHz compared to 2.1 GHz allows us to more readily see the widths of these microcomponents, which reveals that PSR B1937+21 is "on" for at least a large majority of its rotation, a conclusion that would be contrary to that drawn from data that is not sensitive enough to resolve these microcomponents.

The polarization properties of our observations again reflect those previously published. The linearly polarized emission shows the same basic structure as at 2.1 GHz, however we can much more clearly see just how linearly polarized many of the microcomponents are. In contrast, none of the microcomponents show any circular polarization.

### 4.17. *PSR J1944+0907*

#### 4.17.1. *2.1 GHz*

The 2.1 GHz PSR J1944+0907 data have comparatively low SNR, making us unable to report a detailed description of its 2.1 GHz pulse profile. Still, we can see that the 2.1 GHz pulse profile is relatively broad, and includes two components of roughly equal peak intensities. We cannot report any significant detection of linear or circular polarization.

#### 4.17.2. *1.4 GHz*

PSR J1944+0907's 1.4 GHz profile is incredibly complex, showing at least eight distinguishable components, including a low-intensity trailing component. The width of the 1.4 GHz total intensity profile is similar to that of the 2.1 GHz profile, however the leading pulse component is seen to be significantly brighter than any other component in the profile.

The linear polarization profile is similarly complex; perhaps more so. Indeed, there are eleven distinct components. Although many of the components are apparently coincident with corresponding components in the total intensity profile, the density of profile components makes it difficult to ascribe one component to another (or rather, makes it difficult *not* to), which would call into question the physical relevance of any such ascription.

Unsurprisingly, the circular polarization profile is also very complex, with at least three reversals in the sense of circularly polarized emission. The two brightest components are LCP peaks, although there are many other lower-intensity LCP and RCP components.

#### 4.17.3. *430 MHz*

Champion et al. (2005) shows that PSR J1944+0907's 430 MHz total intensity profile consists of two main components that, together, form a profile that looks very similar to the 1.4 GHz profile, albeit with far less complexity. Our data agrees with this, however, we are able to detect a tail on the trailing edge of the profile.

As with the 1.4 GHz data, PSR J1944+0907 shows linearly polarized emission throughout virtually the entire total intensity pulse at 430 MHz, although again there is far less structure apparent to the linear polarization profile. We see at least 3 components, one of which aligns with the brightest total intensity component.

The circularly polarization profile looks very different at 430 MHz than it does at 1.4 GHz in that the emission is almost entirely RCP with one (or possibly two) change in the sense of circularly polarized emission.

### 4.18. *PSR J1949+3106*

#### 4.18.1. *2.1 GHz*

Multi-frequency polarimetry for PSR J1949+3106 was taken upon its discovery (Deneva et al. 2012), and what was found was that PSR J1949+3106 does not show much polarization at all at both 820 MHz and 1.4 GHz. This is also true for PSR J1949+3106 at 2.1 GHz, although there is some linearly polarized emission spanning the duration of the total intensity emission. There is no detectable circularly polarized emission.

Deneva et al. (2012) reported a total intensity profile with two bright peaks, and that the first component has a steeper spectral index than the last. They report that these characteristics are consistent over both observation frequencies (820 MHz and 1.4 GHz), and we find that these characteristics are consistent with the 2.1 GHz emission.

#### 4.18.2. *1.4 GHz*

Our 1.4 GHz data are consistent with the profile reported in Deneva et al. (2012), although the improved sensitivity shows that the linear polarization profile consists of at least 3 components, the brightest on the leading edge and another bright component on the trailing edge. The circularly polarized emission is weak, but shows a RCP peak coinciding with the trailing total intensity peak followed by a reversal of the sense of circular polarization, then a LCP peak coinciding with the trailing linear polarization peak. This phenomenology contrasts with the emission seen at 2.1 GHz, where there the linearly and circularly polarized emission was much weaker, as was the peak intensity of the trailing component of the total intensity profile relative to the leading component.

### 4.19. *PSR B1953+29*

#### 4.19.1. *1.4 GHz*

The 1.4 GHz profile for PSR B1953+29 has been well studied (Boriakoff et al. (1983), Kramer et al. (1998), and Gonzalez et al. (2011)), and includes a bright, multi-component pulse preceded by a low-intensity component. The linear polarization profile shares much of the same shape as the total intensity profile as every total intensity component has a corresponding linearly polarized component. The same can be said of the circularly polarized emission, with the exception of the first two total intensity components. The circular polarization profile starts as LCP, then reverses sense and finished the profile as RCP. Interesting to note is the fact that even though the linearly and circularly polarized emission share very similar shapes, their components appear to be offset from each other by a phase of $\sim 0.005$.

### 4.20. *PSR J1955+2527*

#### 4.20.1. *1.4 GHz*

As with PSR J1949+3106, polarimetry data for PSR J1955+2527 was taken upon its discovery (Deneva et al. 2012), although the data did not have a high enough SNR to detect any linear or circular polarization. Our data are sensitive enough to detect polarized emission, and it is indeed at a low enough level to be consistent with no detection at a lower SNR. Both the linear and circular polarization profiles show one broad pulse spanning nearly the entire duration of the total intensity pulse. As for the total intensity profile, it is relatively simple, though there does appear to be a bump on the leading edge. This is interesting because Deneva et al. (2012) proffered the possibility of an "unresolved bump on the leading edge of the main pulse and/or a slight bump at the very top of the pulse" as a possible explanation for non-gaussian characteristics of the pulse profile.

#### 4.20.2. *430 MHz*

### 4.21. *PSR J2017+0603*

#### 4.21.1. *2.1 GHz*

The 2.1 GHz total intensity profile for PSR J2017+0603 is broadly comprised of two parts, both consisting of multiple components themselves. The first of these two parts is preceded by a tail of emission and is followed by a bridge of emission that connects the two major parts of the total intensity emission. The linear polarization profile is similar in many ways to the total intensity profile, as the components of the linear polarization profile appear to be able to be mapped to components of the total intensity profile bijectively. The relative intensities of these components, however, are not related to the relative intensities of their total intensity counterparts in general. The circular polarization profile is considerably simpler and weaker than the total intensity and linear polarization profiles, as there are only two resolvable RCP components, each coincident with brightest two components of the total intensity profile.

#### 4.21.2. *1.4 GHz*

The 1.4 GHz total intensity profile looks very similar to the 2.1 GHz total intensity profile: we see two main parts, connected by a bridge of emission and preceded by a tail of emission. The exceptional SNR of PSR J2017+0603's 1.4 GHz profile reveals that as complex as the 2.1 GHz emission is, the 1.4 GHz emission shows even more complexity. The linearly polarized emission varies from almost perfectly tracing out the total intensity profile with 100% of the emission being linearly polarized, to straying significantly from the total intensity profile, showing almost no polarized emission. There appears to be no detectable circularly polarized emission, save for two very weak RCP pulses aligning with the most intense components of each of the main parts of the total intensity profile.

### 4.21.3. *430 MHz*

As with the 2.1 and 1.4 GHz emission, we again see two parts to the total intensity emission at 430 MHz, however (perhaps due to the low SNR), we see neither a leading tail nor a bridge of emission. Moreover, and again perhaps due in part to the low SNR, the total intensity profile shows significantly less structure at 430 MHz than at higher frequencies. Further differences between the 430 MHz profile and the higher frequency profiles, and ones that are not due to a low SNR, are seen in the linear polarization profile. Linearly polarized emission, which at times constituted nearly 100% of the total 2.1 and 1.4 GHz emission, is not significantly detected in any part of the pulse profile. We do, however, detect a RCP component of emission that spans much of the latter part of the total intensity profile.

### 4.22. *PSR J2019+2425*

#### 4.22.1. *1.4 GHz*

The 1.4 GHz profile for PSR J2019+2425 consists of 3 parts: a main, dual-component part flanked by two lower-intensity components (Nice et al. 2001). The two flanking components, which are very similar in total intensity, yet are polar opposites in linear polarization. The component preceding the main pulse shows no detectable linear polarization, whereas the component following the main pulse appears to be 100% linearly polarized. This morphology persists at low frequencies (Nice et al. 1993). The main pulse itself shows two linear polarization components, one on the leading edge of the pulse, and one on the trailing edge of the pulse. Interestingly, the local minimum of the linear intensity profile between these two components is coincident with one of the peaks of total intensity emission. There is only one weak LCP component to the circularly polarized emission. This component is coincident with the aforementioned linear polarization local minimum.

### 4.23. *PSR J2033+1734*

#### 4.23.1. *1.4 GHz*

PSR J2033+1734 was discovered with the AO at 430 MHz, and its profile was reported to be a bright pulse followed by a long tail (Ray et al. 1996). This phenomenology also accurately describes PSR J2033+1734's 1.4 GHz profile, although the "tail" appears to be more accurately described as a flat, shelf-like feature, followed by a more conventional, gaussian component. The profile shows significant linear polarization, which mostly occurs on the leading edge of the pulse, leaving the trailing components unpolarized. There is a moderate LCP leak, followed by a sense reversal coinciding with the total intensity peak, leading to a more intense RCP peak. As with the linear polarization, there is no detectable circular polarization associated with the trailing components of the total intensity emission.

#### 4.23.2. *430 MHz*

As described above, PSR J2033+1734 was reported upon its discovered with the AO at 430 MHz to have a profile that consists of a bright pulse followed by a long tail (Ray et al. 1996). We report this same general structure, but our coherently dedispersed data show the pulse is much narrower than initially reported, and that the long tail is inherent to the pulsar's emission, rather than being caused by external processes such as propagation through the interstellar medium. The linear polarization profile is much simpler at 430 MHz than at 1.4 GHz, showing one component roughly aligning with the total intensity main pulse. The linear polarization at 430 MHz is also relatively weak, as it has a linear polarization fraction of 0.03, which contrasts with the linear polarization detected at 1.4 GHz, which had a linear polarization fraction of 0.27. The circular polarization profile is similarly simple, as there is again only one detectable component: a RCP pulse which follows the main pulse by a phase of $\sim 0.01$.

### 4.24. *PSR J2043+1711*

#### 4.24.1. *1.4 GHz*

Upon its discovery, PSR J2043+1711's 1.4 GHz profile was reported to be "complex, with several pulsed components" (Guillemot et al. 2012). Improved sensitivity shows a profile that certainly confirms that statement. What appeared to be the brightest single component is revealed to be two components, and the trailing component is shown not only to have a steep drop in the emission that was previously unresolved, but also to not, in fact, be the trailing component after all, as a very weak pulse is detected following it. The linear polarization profile shows similar complexity, with the brightest components being almost 100% polarized, while other components in the total intensity profile show no corresponding linear polarization at all. For as intense and complex as the linear polarization profile appears to be, the circular polarization profile is much more simple: one RCP component aligned with the brightest total intensity component and two weak LCP components aligning with other total intensity components.

#### 4.24.2. *430 MHz*

PSR J2043+1711's profile looks very different at 430 MHz than it does at 1.4 GHz. Where at higher frequencies, there was a trailing component connected to the main pulse profile by a weak bridge of emission, both the bridge and the trailing component are not seen at all at 430 MHz. Instead, we see 5 relatively sharp components. The linear polarization profile resembles the total intensity profile, with every total intensity component except the last one showing a corresponding linearly polarized component. The circular polarization profile almost perfectly mirrors the linearly polarized emission in RCP, although the linear polarization is ∼ 5% less intense.

### 4.25. *PSR J2214+3000*

#### 4.25.1. *2.1 GHz*

PSR J2214+3000's 2.1 GHz profile consists primarily of two pulses, the peaks of which are separated by almost exactly half a rotation. The brightest of these is made up of one bright component preceded by a much dimmer component. The other pulse is comprised of three components. Following the brighter pulse, however, we are able to detect another, very weak feature. This feature is very narrow and approximately as bright as the component leading the brightest component in the total intensity profile. The linear polarization profile shows a similar structure to the total intensity profile. The leading component of the brightest pulse is nearly 100% polarized, whereas the following component is much less polarized. The dimmer of the two main pulses in the total intensity profile also shows significant linear polarization, although we are unable to detect linear polarization from its trailing component. There is no detectable circular polarization throughout the pulse profile.

#### 4.25.2. *1.4 GHz*

PSR J2214+3000's 1.4 GHz profile looks very similar to the 2.1 GHz profile (unsurprising, as this general shape appears to persist down to frequencies as low as 820 MHz (Ransom et al. 2011)). Again, we see two bright pulses separated by about half a rotation, and again, we are able to detect a very faint pulse trailing the brightest of the total intensity pulses. At 1.4 GHz, however, we are also able to detect a tail leading this very faint pulse. The linear polarization profile still shows a similar structure to the total intensity profile, and the leading components of both pulses are very highly polarized, however at 1.4 GHz we find that the brightest component in the total intensity profile is much more highly polarized. We also find that the trailing component in the dimmer total intensity pulse does have corresponding linear polarization. We are also able to see that the faint, solitary pulse is very highly polarized. There seems to be a broad RCP pulse and LCP pulse coincident with the brighter and dimmer total intensity pulses, respectively, although the peak of the circularly polarized emission is < 1% of the total intensity peak.

### 4.26. *PSR J2229+2643*

#### 4.26.1. *1.4 GHz*

The first published profiles of PSR J2229+2643 at 1.4 GHz (Camilo et al. 1996b) had significantly lower SNR than the profile published in this work, however, since PSR J2229+2643's 1.4 GHz profile is relatively smooth and broad, the increase in sensitivity serves more of a confirmation of previously published profiles than a revelation of previously hidden details. The linear polarization profile contains two components: one relatively sharp component aligned with the total intensity peak, and another component that has a peak intensity less than a fifth that of the brighter component on the leading edge of the profile. The circular polarization profile shows a RCP peak, followed by a sense reversal, a LCP peak, another sense reversal, and an final RCP peak. The LCP peak appears to be aligned with the total intensity peak.

#### 4.26.2. *430 MHz*

Since the 1.4 GHz profile was relatively smooth and broad, and therefore served as more as a confirmation of previously published profiles, it is perhaps not a surprise that the same is true at 430 MHz. Indeed, the total intensity profile shows the same general shape at 430 MHz as it did at 1.4 GHz, a phenomenology that has already been established (Camilo et al. 1996b). For as little as the total intensity profile changes from 1.4 GHz to 430 MHz, the polarization profiles at 430 MHz are starkly different. We detect no linear polarization, and the only circular polarization we detect is a weak LCP peak that trails the total intensity peak by a phase of ∼ 0.02.

### 4.27. *PSR J2234+0611*

#### 4.27.1. *1.4 GHz*

PSR J2234+0611's 1.4 GHz profile appears to be relatively narrow, however closer inspection reveals that it radiates over a substantial fraction of its rotation. The profile shows many components, however most of the components have peak fluxes that are an order of magnitude or two lower than the peak flux of the profile as a whole. These low-intensity components show a high degree of linear polarization in general, although the trailing component appears to be unpolarized. The circular polarization is weak, and shows two LCP peaks that are coincident with the two strongest peaks in the linear polarization profile.

### 4.28. *PSR J2234+0944*

#### 4.28.1. *2.1 GHz*

The total intensity profile for PSR J2234+0944 at 2.1 GHz appears to be made up primarily of to two separate pulses, however, the high SNR of the data show that the pulses that comprise the profile are in fact connected by a bridge of emission. Interestingly, a tail of emission leads into this bridge from both sides. The profile shows moderate linear polarization, with a dual-component pulse coincident with the larger of the total intensity pulses and a very low-intensity component coincident with the smaller of the total intensity pulses. The circularly polarized emission is simple, with two components aligning roughly with the total intensity pulses. These pulses are both LCP, and therefore PSR J2234+0944 does not display a detectable reversal in the sense of circularly polarized emission at 2.1 GHz.

#### 4.28.2. *1.4 GHz*

The total intensity profile for PSR J2234+0944 at 1.4 GHz is very similar to its 2.1 GHz counterpart in that it also appears to be made up primarily of to two separate pulses that are connected by a bridge of emission, however, we are able to detect microcomponents preceding the total intensity peak. The first of these microcomponents does not appear to be polarized, while the second appears to be 100% linearly polarized. The polarized emission coincident with the high-intensity pulses in the total intensity profile are also similar to the corresponding 2.1 GHz components, however we note that a tail emission precedes the final linearly polarized emission, and the higher SNR of the 1.4 GHz profile reveals a reversal in the sense of circularly polarized emission.

#### 4.28.3. *430 MHz*

Although the 430 MHz profile for PSR J2234+0944 resembles the high-frequency profiles in that it is broadly made up of two pulses, we show that the trailing pulse is stronger at 430 MHz whereas the preceding pulse was stronger at higher frequencies, indicating that the spectrum of the trailing pulse is steeper than the leading pulse. We detect a linearly polarized component coincident with the first total intensity peak as well as very weak RCP and LCP pulses corresponding to the two respective total intensity pulses, with a sense reversal in between.

### 4.29. *PSR J2317+1439*

#### 4.29.1. *1.4 GHz*

PSR J2317+1439's 1.4 GHz profile shows a bright, extremely complex main pulse and a low-intensity postcursor. Kramer et al. (1998) reported that PSR J2317+1439's total intensity pulse consisted of 4 components (not counting the postcursor), but we show that is considerably more complex. They also report the presence of a precursor preceding the main pulse by about 0.4 rotations. This precursor was reported to be about half as bright as the postcursor. As the postcursor is easily detectable in our data, we would expect such a precursor to also be easily detectable. As we do not detect it, we conclude that the precursor reported in Kramer et al. (1998) cannot be attributed to the pulsar's intrinsic emission.

The linear polarization profile also shows significant complexity, with the leading components being much more polarized than the trailing components, with the exception of the postcursor (which is nearly totally polarized). The circular polarization profile shows many components, but interestingly, they are all LCP, meaning that despite the complexity, there is no detectable change in the sense of circular polarization.

#### 4.29.2. *430 MHz*

Our total intensity profile for PSR J2317+1439 agrees well with previously reported 430 MHz profiles (Camilo et al. 1993). The linear polarization profile is significantly simpler at 430 MHz than at 1.4 GHz, as there are only two components: one broad component aligned with the first bright total intensity component, followed by a much weaker, much narrower component. The circular polarization profile, on the other hand, is much more complex at 430 MHz. While there was not one detectable reversal of sense of circular polarization at 1.4 GHz, there are many at 430 MHz.

## 5. RESULTING POLARIMETRIC RESPONSES

Recalling that the METM method was performed on profiles that had already been calibrated with the nominal PR created through the MEM procedure, the PRs shown in Figures 18 and 19 can be viewed as "residual PRs" (that is, corrections to the nominal, MEM-generated PR). Since new PRs were produced spanning a range of epochs, we can use them to describe the stability of the PR of AO's 1.4 and 2.1 GHz receivers.

First, however, it is important to be clear about the limitations of the analysis presented in this paper. The ideal standard profiles to use would be ones that were produced using the MEM procedure. As such profiles were not available to us in this data set, we chose profiles that were consistent with previously published profiles. Inevitably, the standard profiles we chose will not be in perfect agreement with the true profiles of the sources we used to make the PRs. Any difference between the two will cause corresponding deviations that will propagate through the PRs made with the imperfect standard profiles, and since these imperfections will be caused by errors in the PRs[5], we expect the errors introduced by using the METM procedure to be frequency-dependent.

---

[5] We assert this due to the stability that PSRs J1713+0747 and B1937+21's polarization profiles display in a number of published analyses over a wide range of epochs.

Crucially, though, it is important to recall that using only the PR produced by the MEM procedure yielded profiles for PSRs J1713+0747 and B1937+21 that were not consistent epoch to epoch, whereas using the PRs created by the METM procedure produces profiles that are consistent with a profile that varies slightly from the ideal standard profile. As the latter is undoubtedly preferable to the former, and as the (albeit imperfect) standard profile does indeed appear to agree with published profiles, we can assume that errors caused by these considerations are small, especially compared to the errors that would have arisen if the METM procedure had not been implemented.

With these considerations in mind, we turn to the residual PRs themselves. In light of the previous discussion, and becuase the METM procedure treats the entire receiver chain as a black box, we do not aim to determine what specific components of the telescope receiver could provide an explanation for any behavior shown in the residual PRs. We also note that any errors in the standard profile will bias the resultant PRs in the same way. Therefore, while those errors will affect the residual PRs in an absolute sense, the time-variability of the PRs will be largely unaffected.

We quantify the variability of the parameters that make up the residual PRs in two ways. First, for each parameter, we calculate the reduced $\chi^2$ for that parameter across all frequency channels on a given epoch. Assuming the standard profile errors are small, this describes how much the given parameter from the residual PR on that epoch varies from the nominal parameter from the residual PR produced by the MEM procedure. Second, for each parameter, we calculate the reduced $\chi^2$ for that parameter across all epochs in a given frequency channel. This describes how much the given parameter varies in that frequency channel on an epoch-by-epoch basis. Again, this variability will be largely unaffected by standard profile errors.

We find that the resulting reduced $\chi^2$ values, both while considering the variation of parameters on a specific epoch, as well as in a specific frequency channel, are extremely large. While high reduced $\chi^2$ can be interpreted as an indication that the errors on parameters are underestimated, it also can serve as quantitative evidence that the residual PR (and therefore the PR) of AO is highly variable. We find this description of AO's PR to be consistent with both the 1.4 GHz PRs and the 2.1 GHz PRs.

## 6. DISCUSSION

The profiles presented are among the highest SNR polarization profiles to date, and this sensitivity reveals that the profiles of many millisecond pulsars have much more structure than may appear without careful inspection. That profile components can be hidden by insufficient SNR is not a new revelation, however, the detection of microcomponents in pulse profiles, particularly the microcomponents detected in PSRs B1937+21, J1713+0747, and J2234+0944 detected for the first time, challenges the very notion of "sufficient" SNR.

These microcomponents also complicate processes that aim to rigorously define "on-pulse" and "off-pulse" regions, as many such routines implicitly or explicitly assume that profile bins that have intensities which are "small" compared to the brightest components in the profile must be noise, as assumption which is refuted by the existence of microcomponents. This has implications for any process which requires the specification of "on-pulse" and "off-pulse" regions. Observers who would like to flux calibrate their observations by calculating radiometer noise, for example, may derive a substantially overestimated value for the radiometer noise present in an observation if an "off-pulse region" is not selected with great care with regard to the possible presence of microcomponents. It should be noted that since NANOGrav does not use the radiometer Equation to calibrate their fluxes (Demorest et al. 2013), the reported fluxes of the NANOGrav pulsars are not affected by this effect.

### 6.1. *Position Angle Swings*

Notably, the position angle swings of the polarimetric profiles presented in this analysis are not generally well-described by the RVM shown in Equation 7 (see, for example, the 1.4 GHz profiles of PSRs B1855+09, J1944+0907, J2234+0611, and many more). Consequently, a detailed characterization of the pulsar's spin geometry would necessitate a more complex model of the PA swing (which we leave for future analysis). In lieu of such an analysis, the profiles themselves can be used to make simple statements about a pulsar's spin geometry (see, for example the discussion in Section 4.11.1 regarding whether or not PSR B1855+09 is an orthogonal rotator) or magnetosphere (by using the period-width relationship to constrain the pulsar's beam width). Care must be taken in these cases, however, as the existence of microcomponents can drastically change the interpretation of the pulse profiles and should therefore be taken into consideration when defining quanities such as "pulse width" and "duty cycle".

The PA swings also show the existence of orthogonal polarization modes (OPMs) for a number of sources in this dataset, including PSRs J1713+0747, J1944+0907, B1953+29, J2017+0603, and J2234+0611, as well as non-orthogonally polarized modes of emission (see, for example PSR J0023+0923 at 1.4 GHz). We note that these modes generally persist over different observing frequencies (see, for example, PSRs J1911+1347 and B1937+21 at 2.1 and 1.4 GHz). While the OPMs for these pulsars coincide with either LCP emission (as in the cases of PSRs J1713+0747, J1944+0907, and B1953+29) or no circularly polarized emis-sion (as in the cases of PSRs J2017+0603 and J2234+0611), the sign of the circular polarization does not generally change when returning to the dominant mode of emission. We cannot therefore ascribe a handedness of circular polarization to the mode of emission. This is consistent with some studies of the relationship between PA and circular polarization handedness (see Stinebring et al. (1984), for example), but not others (for instance, Cordes et al. (1978)). We note that while the most detailed studies of the PA-circular polarization handedness relation come from statistical anaylses of pulsar single pulses, our analysis comes from integrated profiles. Still, other analyses of integrated profiles (such as Weisberg et al. (2004)) show further examples of sources that do not display a change of circular polariation handedness with a change of the dominant mode of emission.

Young pulsars have been shown to have smaller degrees of polarization with increasing observing frequency (i.e. negative "depolarization indices", see Manchester et al. 1973; Xilouris et al. 1996). We do not see the same general trend in our data. Rather, we see a spread of depolarization indices, with some showing a smaller degree of polarization with increasing observing frequency, similar to young pulsars, and others showing an opposite dependence on observing frequency. This spread in depolarization index was also seen in Dai et al. (2015).

## 6.2. *Interstellar Scattering*

Propagation of the pulsar signal through the interstellar medium will cause scattering, which serves to smear the profiles, but we determine that this effect is negligible for our data taken at 2.1 and 1.4 GHz, as the timescale for this smearing as determined by Levin et al. (2016) is generally $< 1\%$ of the pulsar's spin period at these frequencies with the exception of PSR J1903+0327 (see Section 4.12). It is unclear what percentage of the pulsar's spin period the scattering timescale is at 430 MHz since this will depend on the spectral index of the scattering timescale, which is not precisely known, but using any reasonable estimate of the scattering spectral index to extrapolate measured 1.4 GHz scattering timescales to 430 MHz indicates that scattering is likely to play a role at these low observing frequencies. This will reduce the degree of polarization and flatten the PA swings (Li & Han 2003; Karastergiou 2009). We do see evidence for flatter PA swings at 430 MHz for several pulsars, including PSRs J0023+0923 and J1640+2224.

## 6.3. *Rotation Measures*

We also present RM measurements for all sources at both 1.4 and 2.1 GHz (where applicable) derived independently for each epoch. We note that for the sources where a RM could be determined independently from the 1.4 GHz ob-servations[6], the RM values are largely in agreement with each other. The source with the largest RM discrepancy versus receiver is PSR J1903+0327, whose RMs at 2.1 and at 1.4 GHz differ by $1.5\sigma$. While this discrepancy may be caused by statistical fluctuation, it is interesting to note that PSR J1903+0327 has the highest DM of any pulsar in the data set, and by far the highest DM of any pulsar for which we are able to measure RMs for 1.4 and 2.1 GHz independently, which means that there is more interstellar material between Earth and the pulsar. As interstellar scattering is a frequency-dependent phenomenon, the radio waves scattered by the ISM will sample a different region of the ISM at different observing frequencies (Cordes et al. 2016). It is possible that the region sampled at 1.4 GHz includes interstellar components and magnetic fields not sampled at 2.1 GHz, thereby changing the properties that govern Faraday rotation, and thus, the observed RM. We stress, though, that we put forth this explanation as another possibility rather than an assertion that it is indeed the cause of the variation in PSR J1903+0327's RM with respect to observing frequency. More observations would be necessary to determine the cause of any such variation.

Using Equation 9, we can calculate the component of the Galactic magnetic field parallel to the line of sight to the source and use it to probe the structure of the Galaxy's magnetic field. We see that the magnetic field strengths are antisymmetric about the Galactic plane, with nearly all the magnetic field strengths above the plane being positive (towards Earth), and nearly all the magnetic field strengths below the plane being negative (away from Earth, see Fig 20). This structure suggests that the magnetic field in the Galactic halo is broadly dipolar, a phenomenology consistent with previous studies (Han et al. 1997; Xu & Han 2014).

There still remains much to be learned from the polarimetric data taken as a result of NANOGrav's timing campaign, and despite this datasets sensitivity, it represents only half of NANOGrav's full polarimetric dataset. The forthcoming release of this full dataset, including the pulsars observed with the Green Bank Telescope, will be accompanied by a more detailed analysis of MSP polarimetry and its implications on MSP magnetospheres, radio emission, and the environments through which that emission propagates.

---

[6] See 3.2 for why this might not be able to be done.

profiles analyzed here. Other contributions are summarized in The NANOGrav Collaboration et al. (2015).

This NANOGrav project received support from the National Science Foundation (NSF) PIRE program award num-

ber 0968296 and NSF Physics Frontier Center award number 1430284. PAG acknowledges Willem van Straten for his invaluable advice. Pulsar research at UBC was supported by NSERC Discovery and Discovery Accelerator Supplement Grants and by the Canadian Institute for Advanced Research.

## REFERENCES

Backer, D. C., & Rankin, J. M. 1980, Astrophys. J. Suppl. Ser., 42, 143

Backer, D. C., Rankin, J. M., & Campbell, D. B. 1976, Nature, 263, 202

Barnard, J. J. 1986, Astrophys. J., 303, 280

Blaskiewicz, M., Cordes, J. M., & Wasserman, I. 1991, Astrophys. J., 370, 643

Boriakoff, V., Buccheri, R., & Fauci, F. 1983, Nature, 304, 417

Camilo, F., Nice, D. J., Shrauner, J. A., & Taylor, J. H. 1996a, Astrophys. J., 469, 819

Camilo, F., Nice, D. J., & Taylor, J. H. 1993, Astrophys. J. Lett., 412, L37

—. 1996b, Astrophys. J., 461, 812

Champion, D. J., Lorimer, D. R., McLaughlin, M. A., et al. 2005, Mon. Not. R. Astron. Soc., 363, 929

Chernyakova, M., Neronov, A., Aharonian, F., Uchiyama, Y., & Takahashi, T. 2009, MNRAS, 397, 2123

Coles, W. A., Kerr, M., Shannon, R. M., et al. 2015, ApJ, 808, 113

Cordes, J. M., & Lazio, T. J. W. 2002, astro-ph/0207156

Cordes, J. M., Rankin, J. M., & Backer, D. C. 1978, Astrophys. J., 223, 961

Cordes, J. M., Shannon, R. M., & Stinebring, D. R. 2016, ApJ, 817, 16

Craig, H. A. 2014, ApJ, 790, 102

Dai, S., Hobbs, G., Manchester, R. N., et al. 2015, MNRAS, 449, 3223

Demorest, P. B., Pennucci, T., Ransom, S. M., Roberts, M. S. E., & Hessels, J. W. T. 2010, Nature, 467, 1081

Demorest, P. B., Ferdman, R. D., Gonzalez, M. E., et al. 2013, ApJ, 762, 94

Deneva, J. S., Freire, P. C. C., Cordes, J. M., et al. 2012, ApJ, 757, 89

Desvignes, G., Caballero, R. N., Lentati, L., et al. 2016, MNRAS, 458, 3341

Dyks, J. 2008, MNRAS, 391, 859

Espinoza, C. M., Guillemot, L., Çelik, Ö., et al. 2013, MNRAS, 430, 571

Everett, J. E., & Weisberg, J. M. 2001, Astrophys. J., 553, 341

Foster, R. S., Cadwell, B. J., Wolszczan, A., & Anderson, S. B. 1995, Astrophys. J., 454, 826

Freire, P. C. C., Wex, N., Esposito-Farèse, G., et al. 2012, MNRAS, 423, 3328

Gonzalez, M. E., Stairs, I. H., Ferdman, R. D., et al. 2011, ApJ, 743, 102

Guillemot, L., Freire, P. C. C., Cognard, I., et al. 2012, MNRAS, 422, 1294

Han, J. L., Manchester, R. N., Berkhuijsen, E. M., & Beck, R. 1997, Astron. Astrophys., 322, 98

Han, J. L., Manchester, R. N., Lyne, A. G., Qiao, G. J., & van Straten, W. 2006, ApJ, 642, 868

Heiles, C. 2002, in Astronomical Society of the Pacific Conference Series, Vol. 278, Single-Dish Radio Astronomy: Techniques and Applications, ed. S. Stanimirovic, D. Altschuler, P. Goldsmith, & C. Salter, 131–152

Jacoby, B. A. 2005, PhD thesis, California Institute of Technology, California, USA

Jacoby, B. A., Bailes, M., Ord, S. M., Knight, H. S., & Hotan, A. W. 2007, ApJ, 656, 408

Johnston, S., & Kerr, M. 2018, MNRAS, 474, 4629

Jones, M., McLaughlin, M., & NANOGrav Timing Group, N. I. G. 2016, in American Astronomical Society Meeting Abstracts, Vol. 227, American Astronomical Society Meeting Abstracts, 435.04

Jones, M. L., McLaughlin, M. A., Lam, M. T., et al. 2017, ApJ, 841, 125

Karastergiou, A. 2009, MNRAS, 392, L60

Komesaroff, M. M. 1970, Nature, 225, 612

Kramer, M., Lange, C., Lorimer, D. R., et al. 1999a, Astrophys. J., 526, 957

Kramer, M., Xilouris, K. M., Camilo, F., et al. 1999b, Astrophys. J., 520, 324

Kramer, M., Xilouris, K. M., Lorimer, D. R., et al. 1998, Astrophys. J., 501, 270

Kuz'min, A. D., & Losovskii, B. Y. 1999, Astronomy Letters, 25, 375

Levin, L., McLaughlin, M. A., Jones, G., et al. 2016, ApJ, 818, 166

Lewandowski, W., Wolszczan, A., Feiler, G., Konacki, M., & Sołtysiński, T. 2004, Astrophys. J., 600, 905

Li, X. H., & Han, J. L. 2003, A&A, 410, 253

Lommen, A. N., Zepka, A., Backer, D. C., et al. 2000, Astrophys. J., 545, 1007

Lorimer, D. R., & Kramer, M. 2005, Handbook of Pulsar Astronomy (Cambridge University Press)

Lorimer, D. R., McLaughlin, M. A., Champion, D. J., & Stairs, I. H. 2007, MNRAS, 379, 282

Lorimer, D. R., Faulkner, A. J., Lyne, A. G., et al. 2006, MNRAS, 372, 777

Lynch, R. S., Boyles, J., Ransom, S. M., et al. 2013, ApJ, 763, 81

Manchester, R. N., Taylor, J. H., & Huguenin, G. R. 1973, ApJL, 179, L7

Manchester, R. N., Hobbs, G., Bailes, M., et al. 2013, Publ. Astron. Soc. Australia, 30, e017

Melrose, D. 2004, in IAU Symposium, Vol. 218, Young Neutron Stars and Their Environments, ed. F. Camilo & B. M. Gaensler, 349

Navarro, J., Manchester, R. N., Sandhu, J. S., Kulkarni, S. R., & Bailes, M. 1997, Astrophys. J., 486, 1019

Nice, D. J., Splaver, E. M., & Stairs, I. H. 2001, Astrophys. J., 549, 516

Nice, D. J., Taylor, J. H., & Fruchter, A. S. 1993, Astrophys. J. Lett., 402, L49

Ord, S. M., van Straten, W., Hotan, A. W., & Bailes, M. 2004, MNRAS, 352, 804

Pennucci, T. T. 2015, PhD thesis, University of Virginia

Radhakrishnan, V. 1969, Proceedings of the Astronomical Society of Australia, 1, 254

Radhakrishnan, V., & Cooke, D. J. 1969, Astrophys. Lett., 3, 225

Rankin, J. M. 1983, Astrophys. J., 274, 333

—. 1990, Astrophys. J., 352, 247

Ransom, S. M., Ray, P. S., Camilo, F., et al. 2011, ApJL, 727, L16

Ray, P. S., Thorsett, S. E., Jenet, F. A., et al. 1996, Astrophys. J., 470, 1103

Segelstein, D. J., Rawley, L. A., Stinebring, D. R., Fruchter, A. S., & Taylor, J. H. 1986, Nature, 322, 714

Stairs, I. H., Thorsett, S. E., & Camilo, F. 1999, Astrophys. J. Suppl. Ser., 123, 627

Stairs, I. H., Faulkner, A. J., Lyne, A. G., et al. 2005, Astrophys. J., 632, 1060

Stinebring, D. R., Cordes, J. M., Rankin, J. M., Weisberg, J. M., & Boriakoff, V. 1984, Astrophys. J. Suppl. Ser., 55, 247

The NANOGrav Collaboration, Arzoumanian, Z., Brazier, A., et al. 2015, ApJ, 813, 65

Thorsett, S. E., & Stinebring, D. R. 1990, ApJ, 361, 644

van Straten, W. 2004, ApJS, 152, 129

—. 2013, ApJS, 204, 13

Weisberg, J. M., Cordes, J. M., Kuan, B., et al. 2004, Astrophys. J. Suppl. Ser., 150, 317

Xilouris, K. M., Kramer, M., Jessner, A., et al. 1998, Astrophys. J., 501, 286

Xilouris, K. M., Kramer, M., Jessner, A., Wielebinski, R., & Timofeev, V. 1996, Astron. Astrophys., 309, 481

Xu, J., & Han, J.-L. 2014, Research in Astronomy and Astrophysics, 14, 942

Yan, W. M., Manchester, R. N., van Straten, W., et al. 2011, MNRAS, 414, 2087

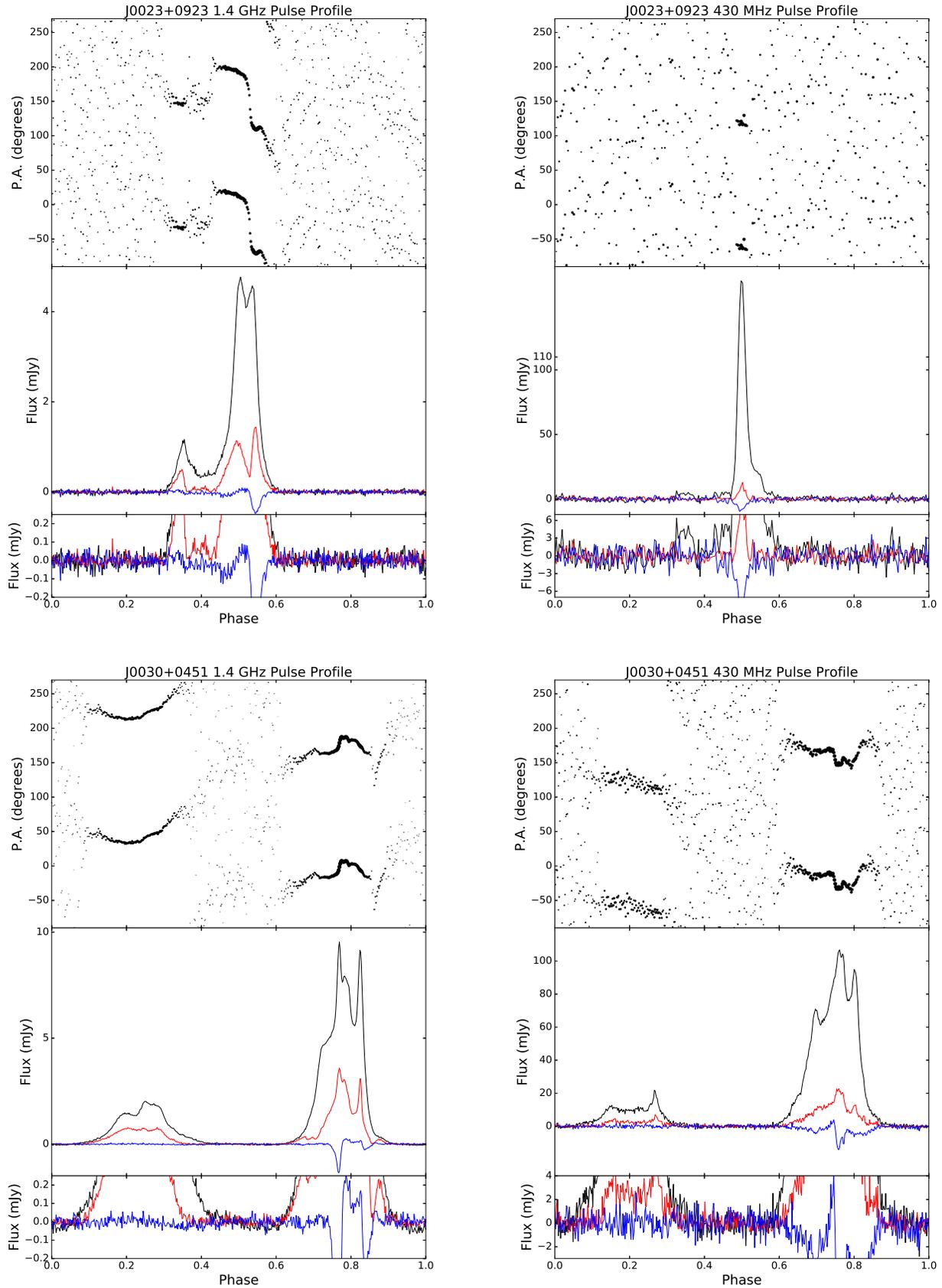

**Figure 1**: Polarization profile of PSR J0023+0923 at 1.4 GHz and 430 MHz, and PSR J0030+0451 at 1.4 GHz and 430 MHz. The top panel of each subplot shows Position Angle (P.A.) in degrees, which are plotted twice for clarity. The middle panel shows the full polarization profile, that is, intensity versus pulsar spin phase. The profile is roughly aligned with the center of the on-pulse region. Total intensity is plotted in black, linear polarization in red and circular polarization in blue. The bottom panel is the same as the middle panel, except zoomed vertically to show any possible microcomponents in more detail. All panels are phase-aligned.

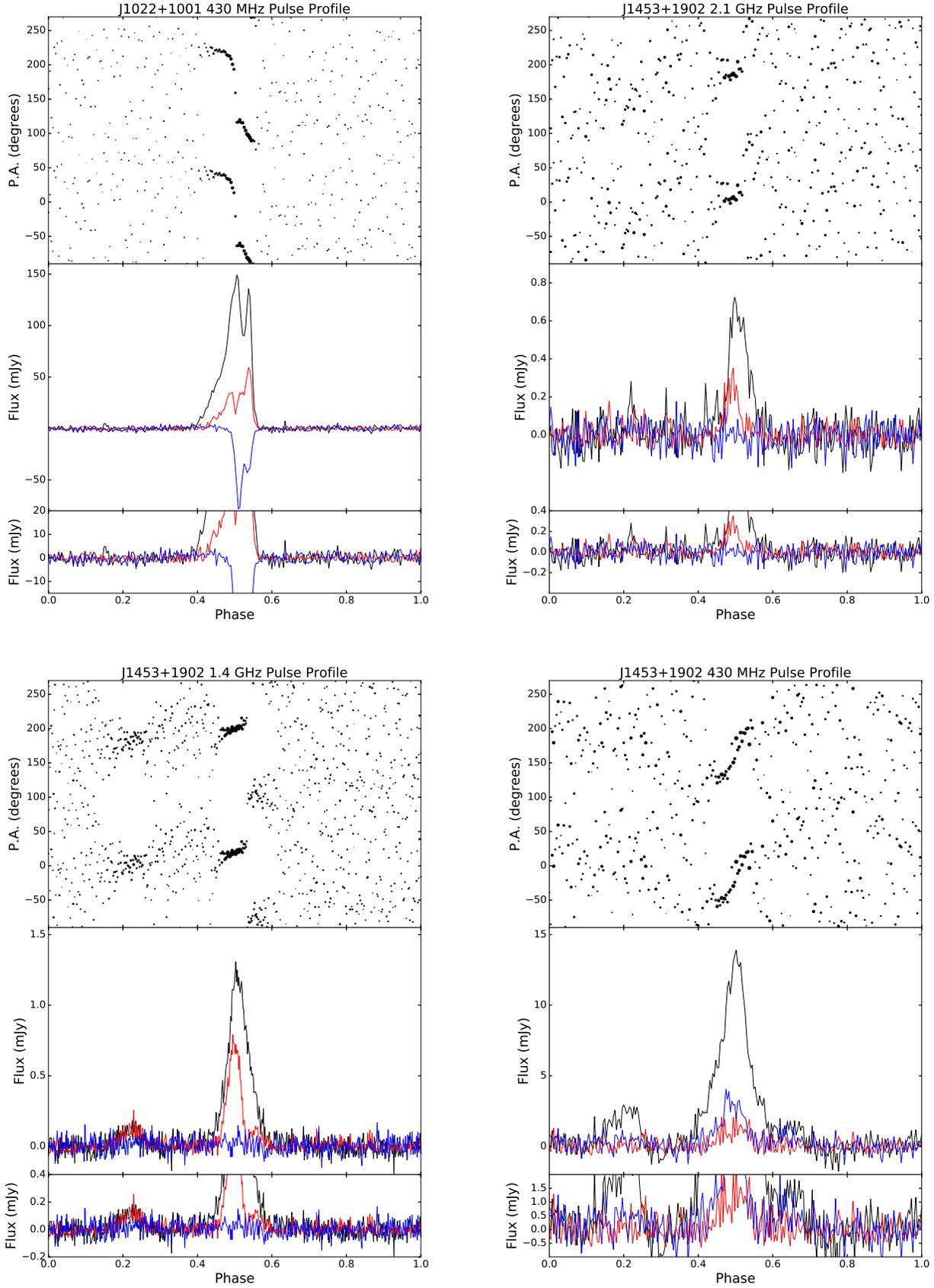

**Figure 2**: Same as Figure 1, for PSR J1022+1001 at 430 MHz and PSR J1453+1902 at 2.1 GHz, 1.4 GHz, and 430 MHz.

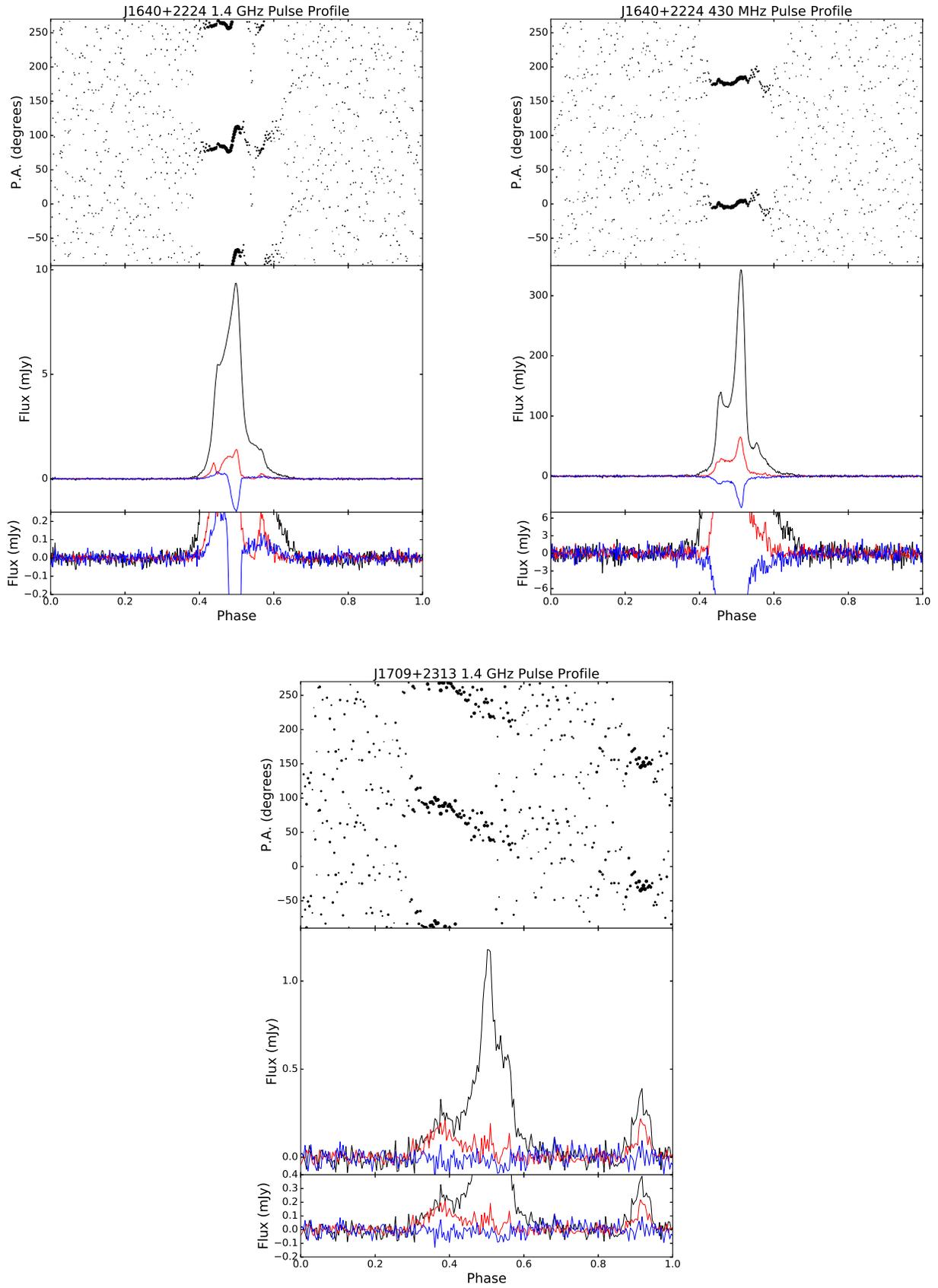

**Figure 3**: Same as Figure 1, for PSR J1640+2224 at 1.4 GHz and 430 MHz, and PSR J1709+2313 at 1.4 GHz.

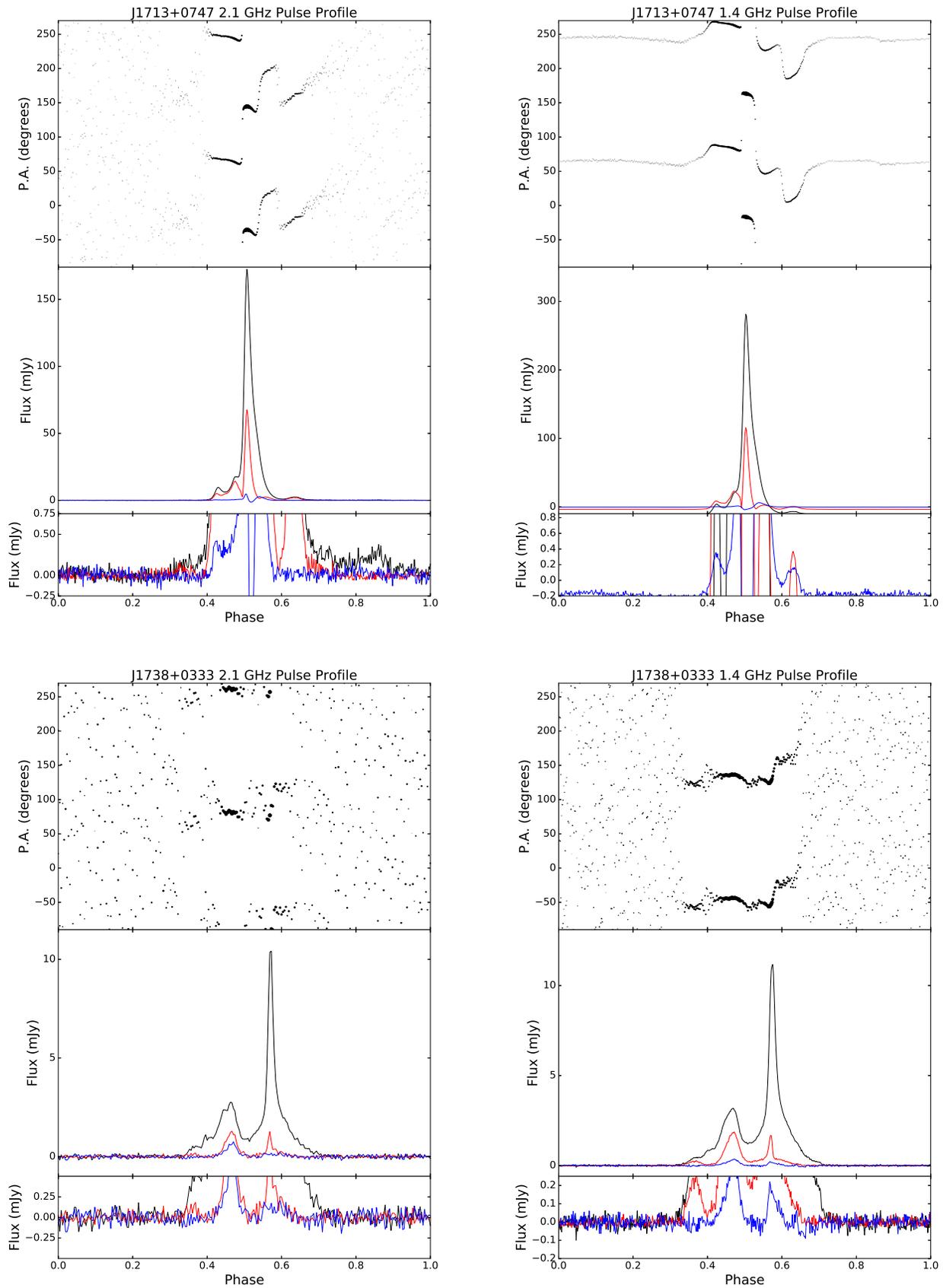

**Figure 4**: Same as Figure 1, for PSR J1713+0747 at 2.1 GHz and 1.4 GHz and PSR J1738+0333 at 2.1 and 1.4 GHz.

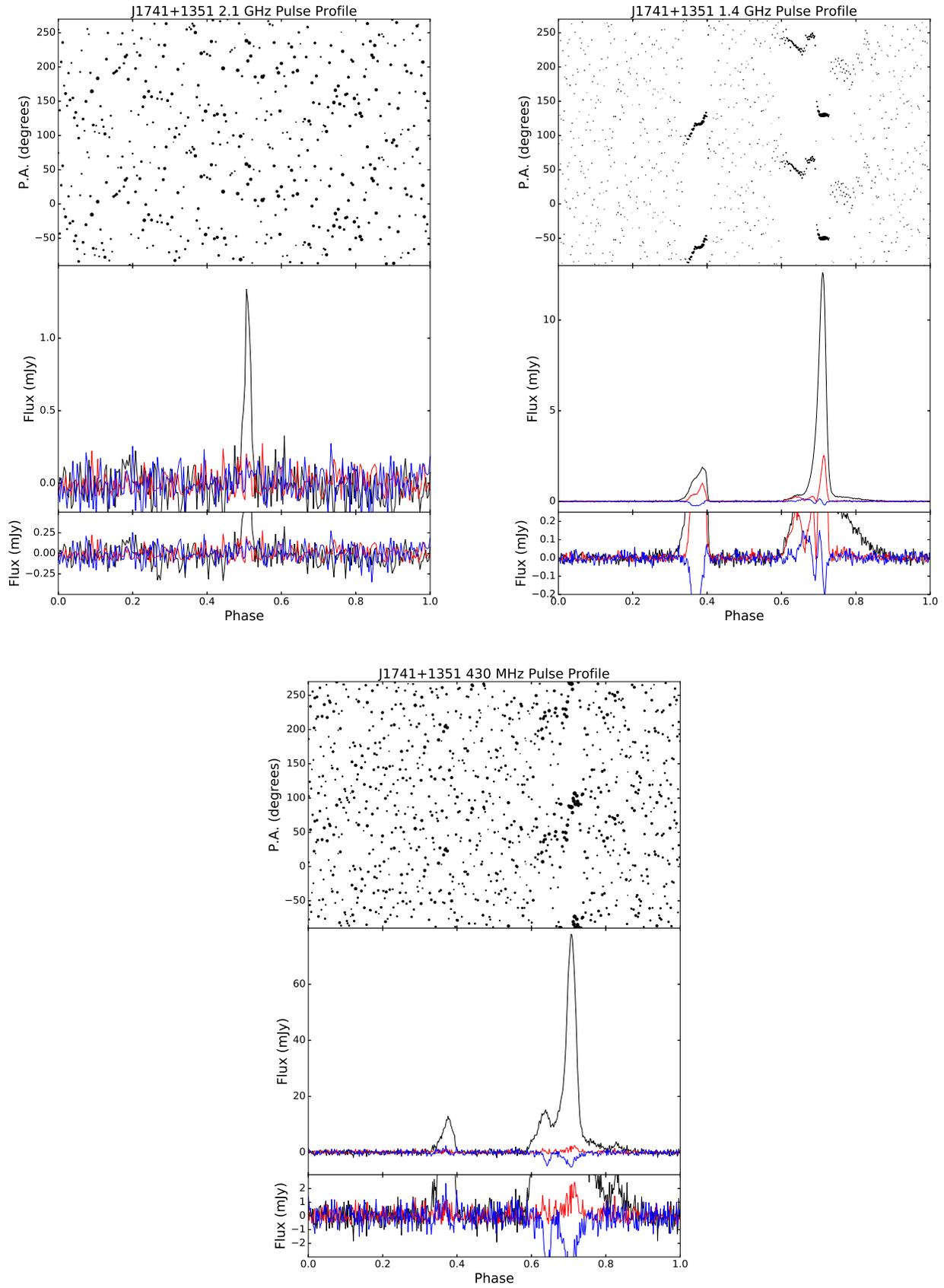

**Figure 5**: Same as Figure 1, for PSR J1741+1351 at 2.1 GHz, 1.4 GHz, and 430 MHz.

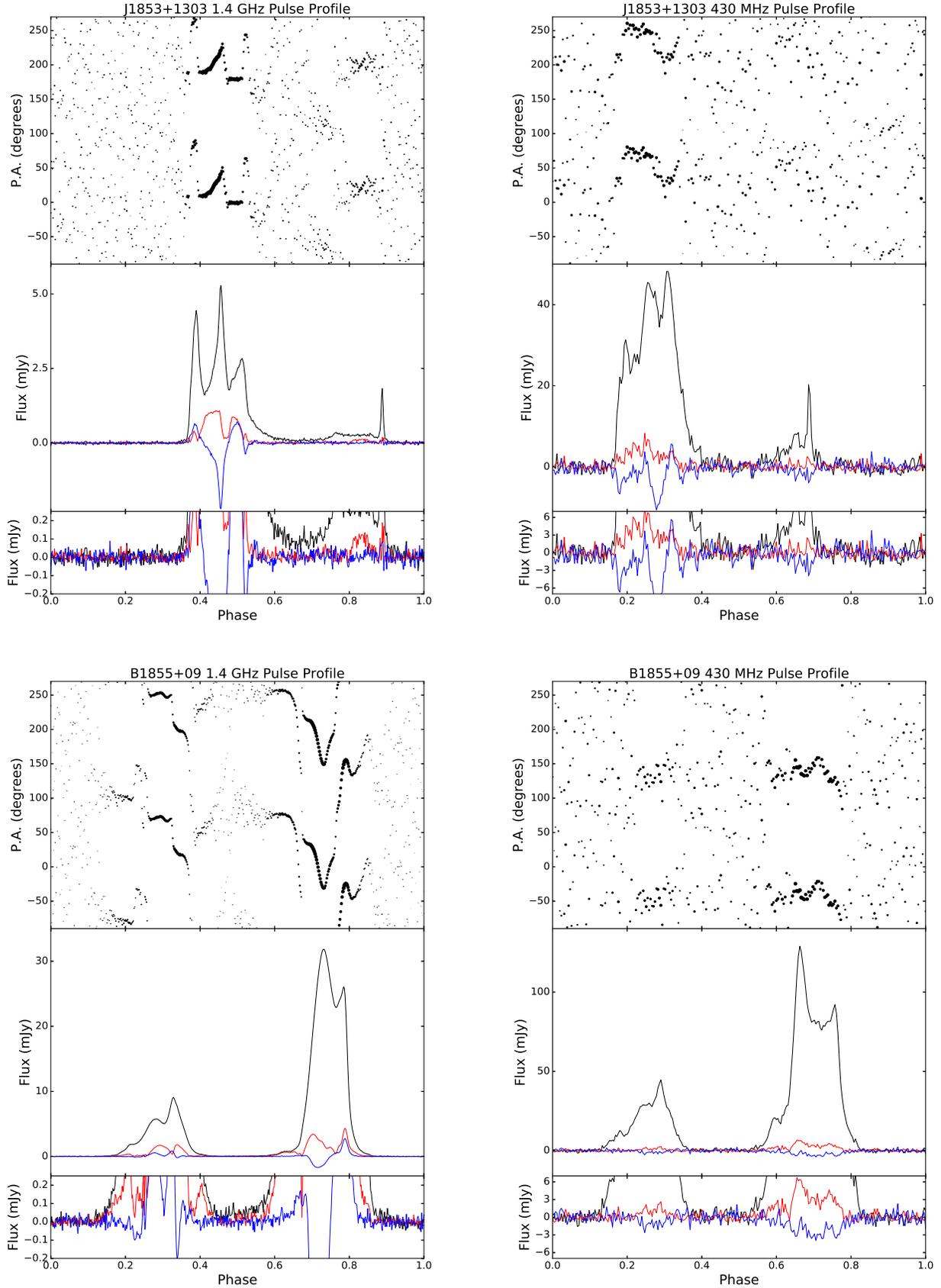

**Figure 6**: Same as Figure 1, for PSR J1853+1303 at 1.4 GHz and 430 MHz, and PSR B1855+09 at 1.4 GHz and 430 MHz.

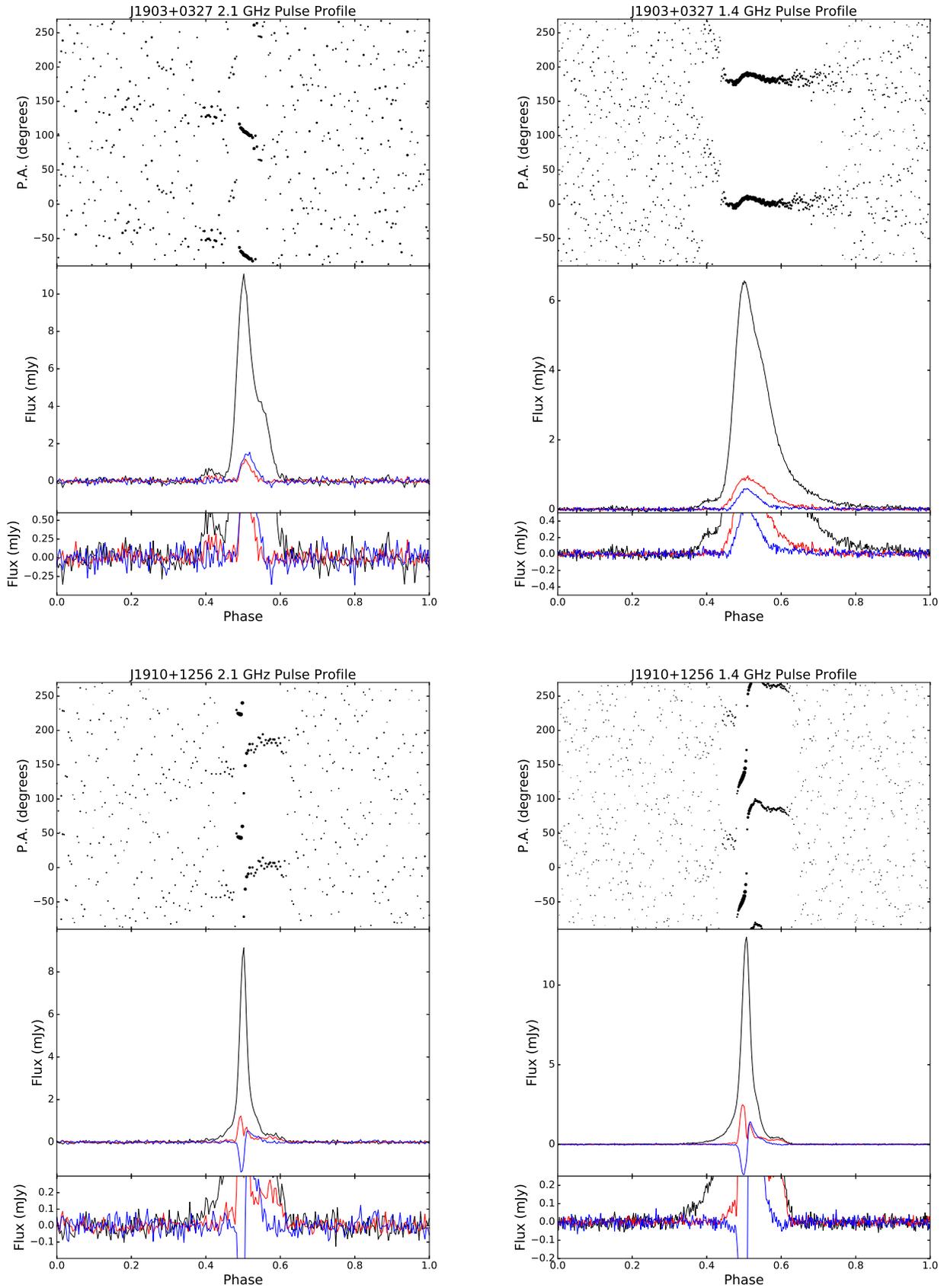

**Figure 7**: Same as Figure 1, for PSR J1903+0327 at 2.1 and 1.4 GHz, and PSR J1910+2156 at 2.1 and 1.4 GHz.

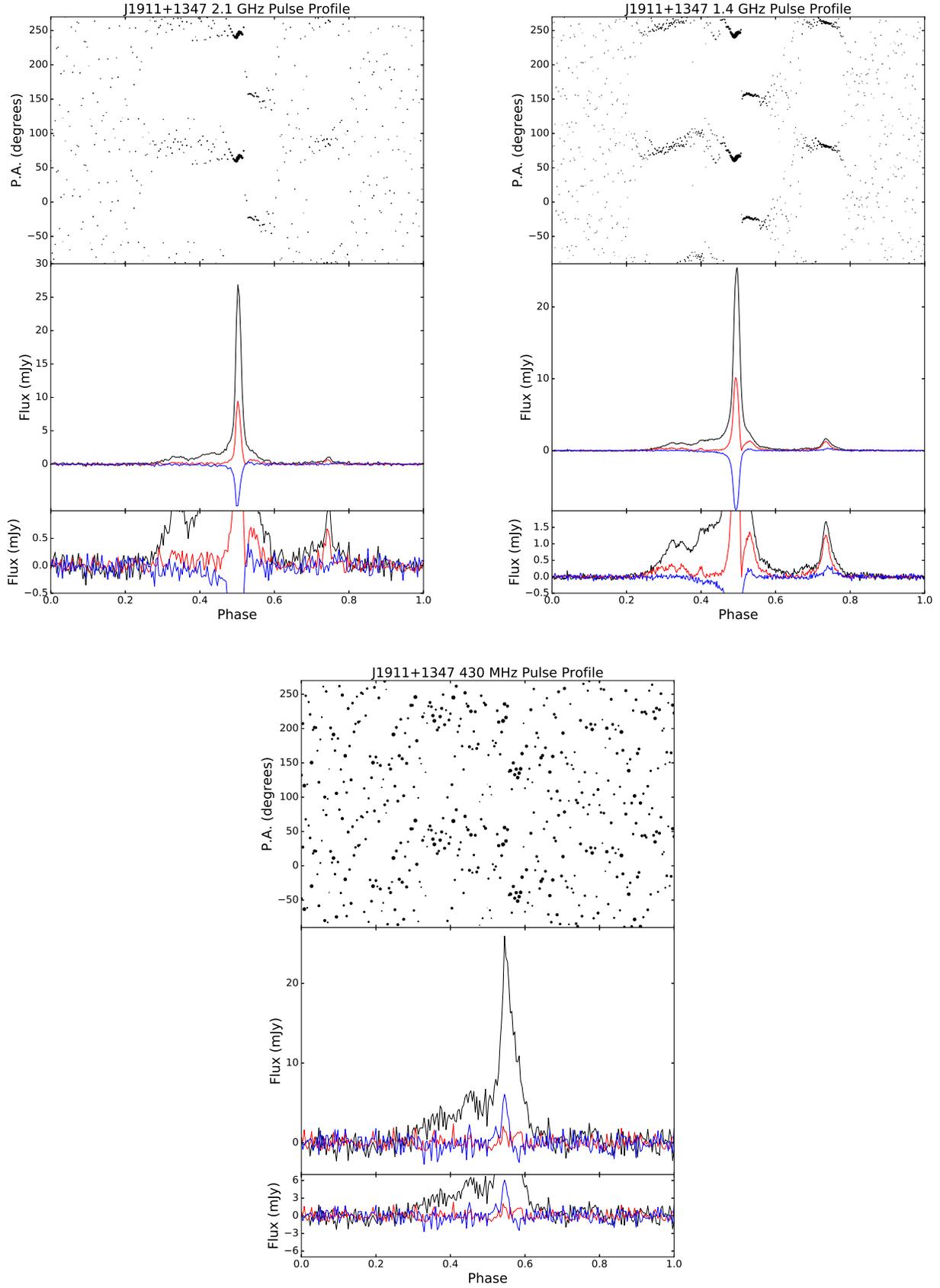

**Figure 8**: Same as Figure 1, for PSR J1911+1347 at 2.1 GHz, 1.4 GHz, and 430 MHz.

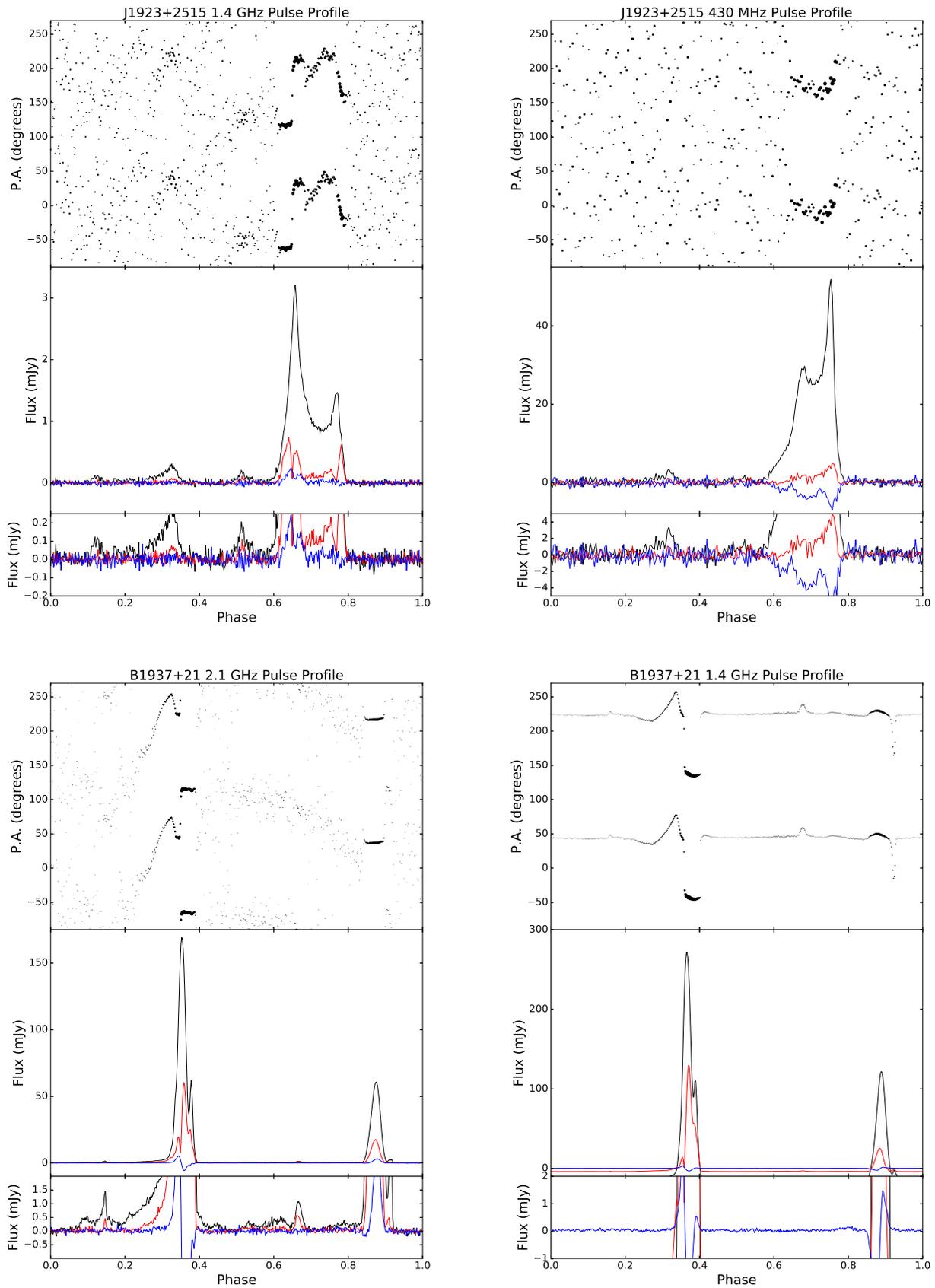

**Figure 9**: Same as Figure 1, for PSR J1923+2515 at 1.4 GHz and 430 MHz, and PSR B1937+21 at 2.1 and 1.4 GHz.

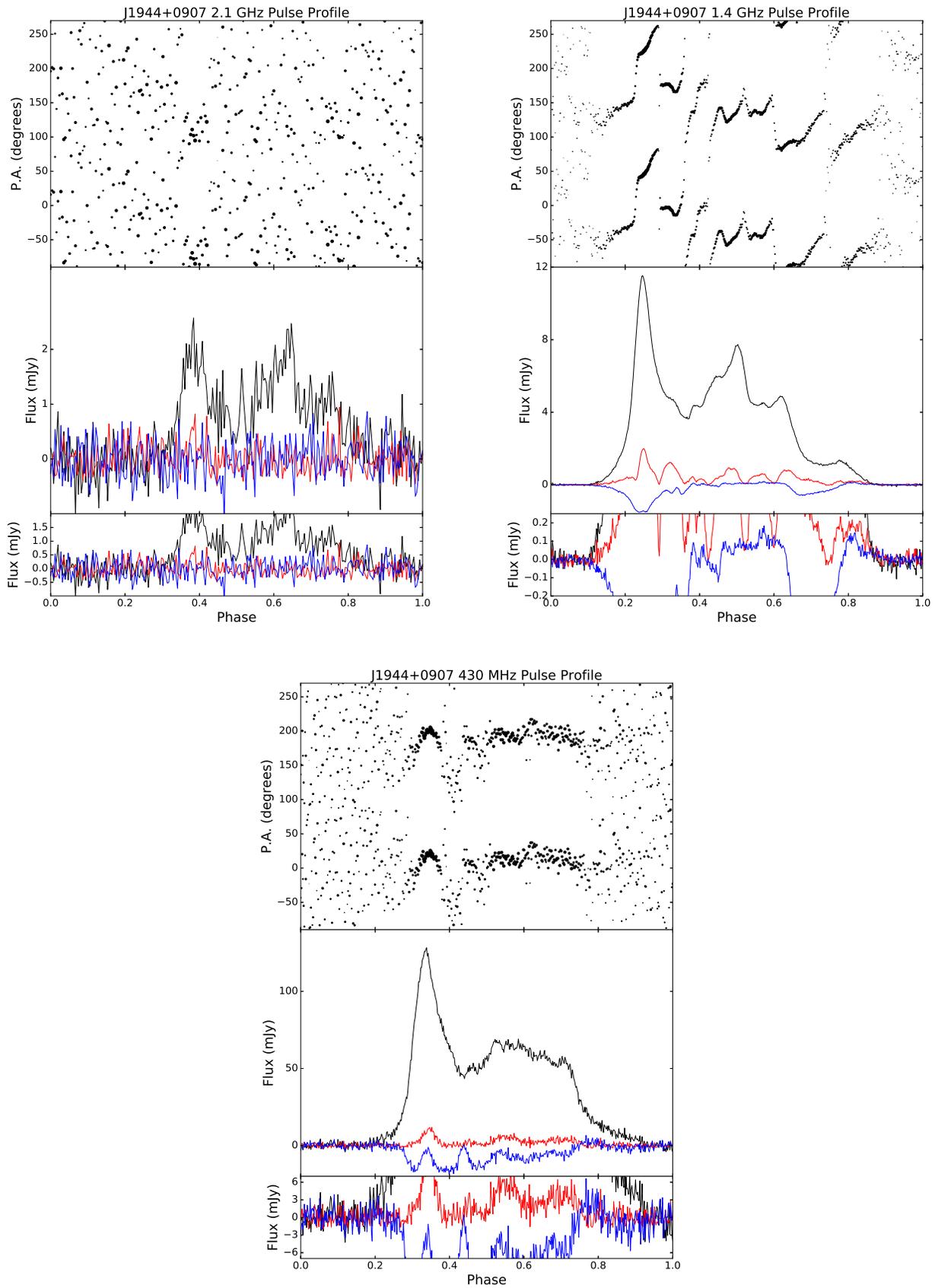

**Figure 10**: Same as Figure 1, for PSR J1944+0907 at 2.1 GHz, 1.4 GHz, and 430 MHz.

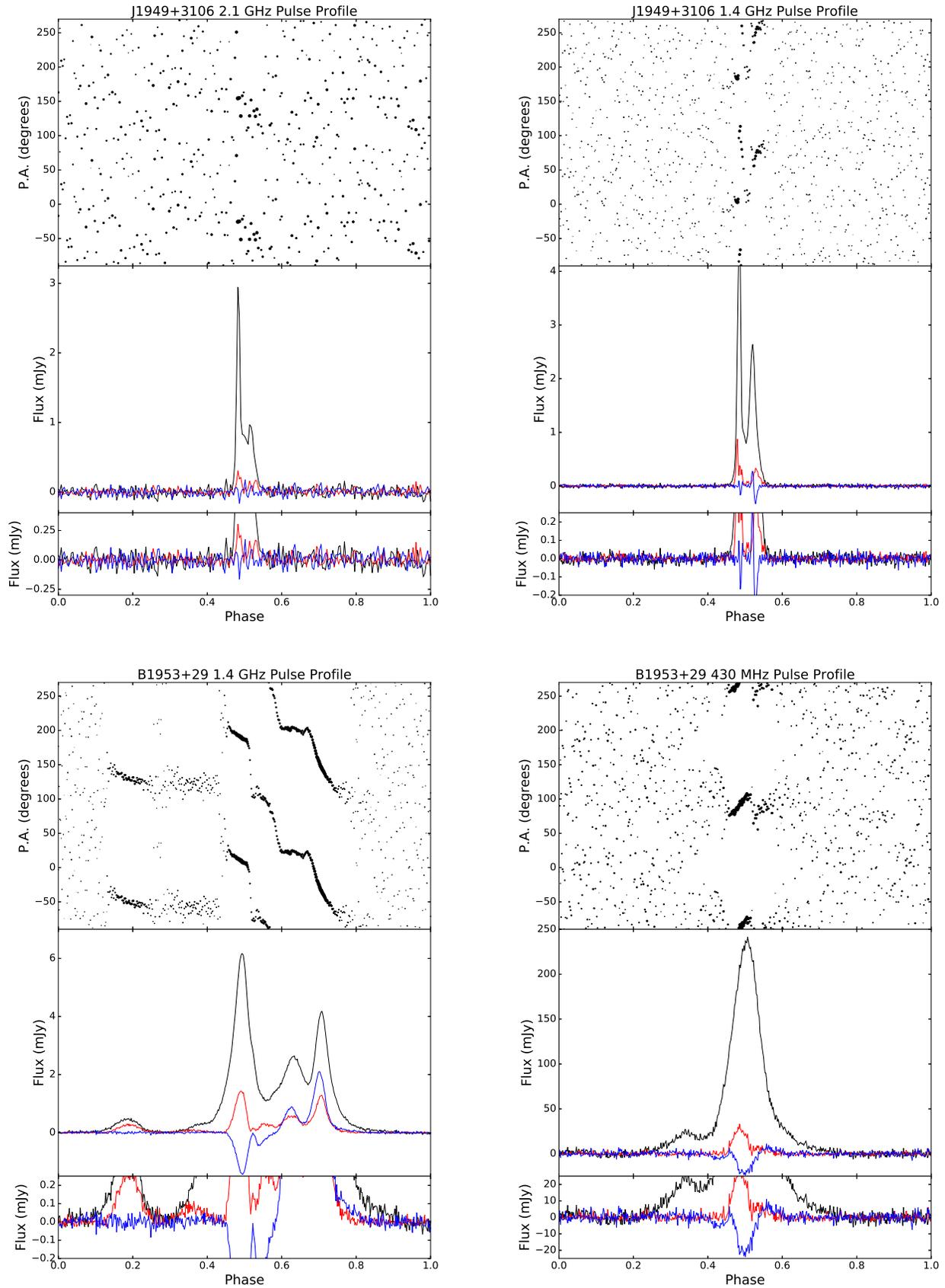

**Figure 11**: Same as Figure 1, for PSR J1949+3106 at 2.1 and 1.4 GHz, and PSR B1953+29 at 1.4 GHz and 430 MHz.

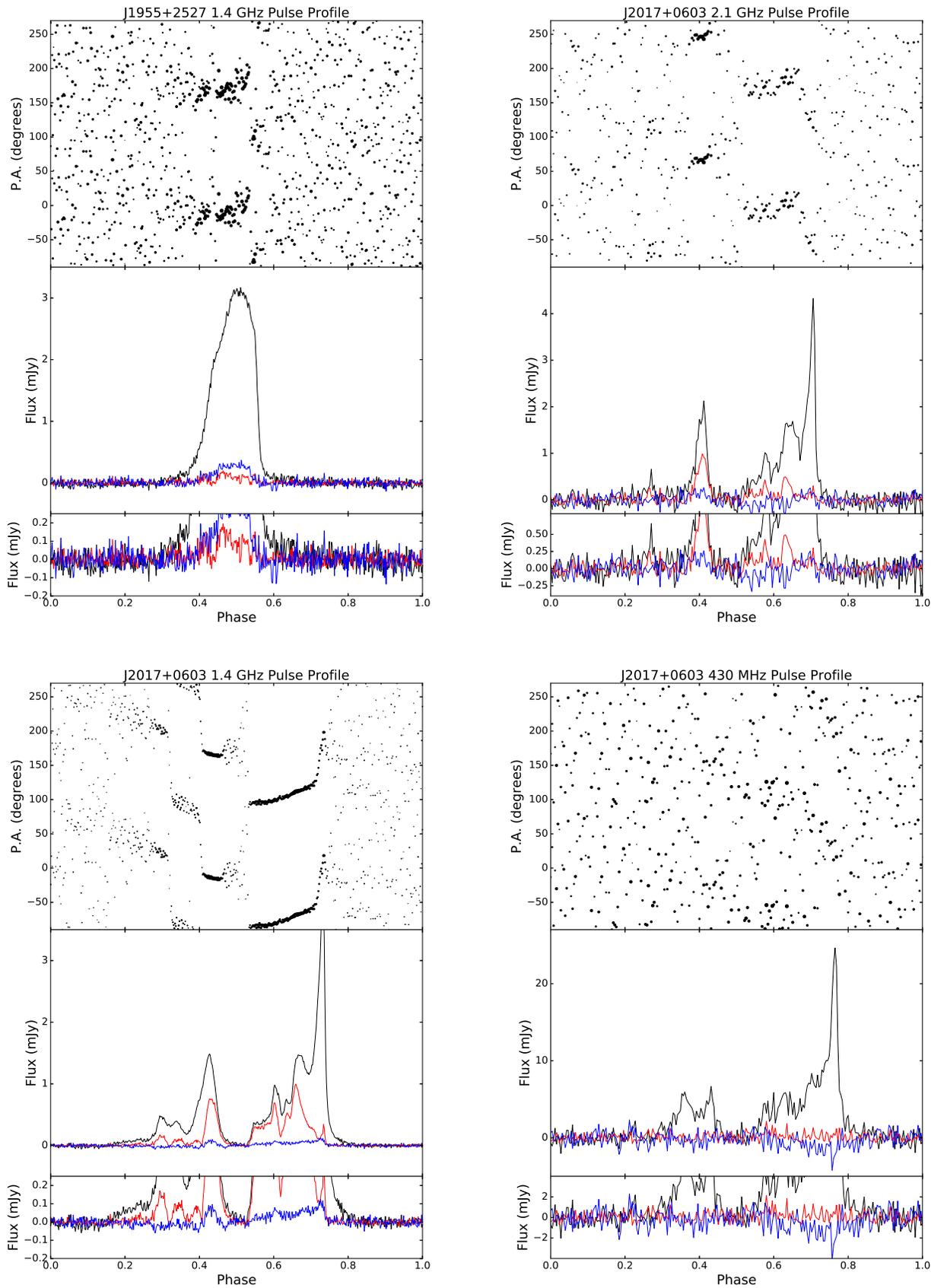

**Figure 12**: Same as Figure 1, for PSR J1955+2527 at 1.4 GHz and PSR J2017+0603 at 2.1 GHz, 1.4 GHz, and 430 MHz.

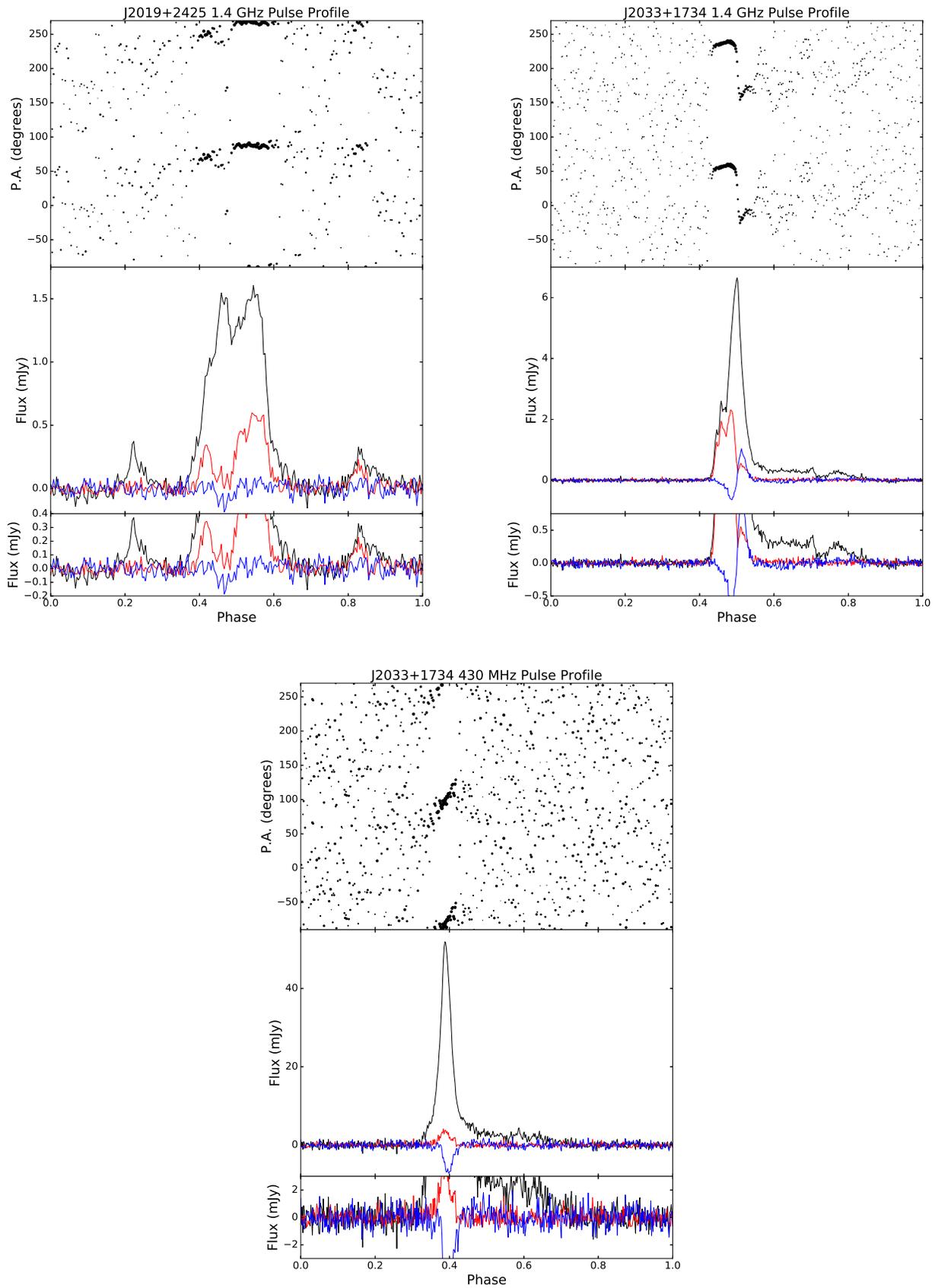

**Figure 13**: Same as Figure 1, for PSR J2019+2425 at 1.4 GHz, PSR J2033+1734 at 1.4 GHz and 430 MHz.

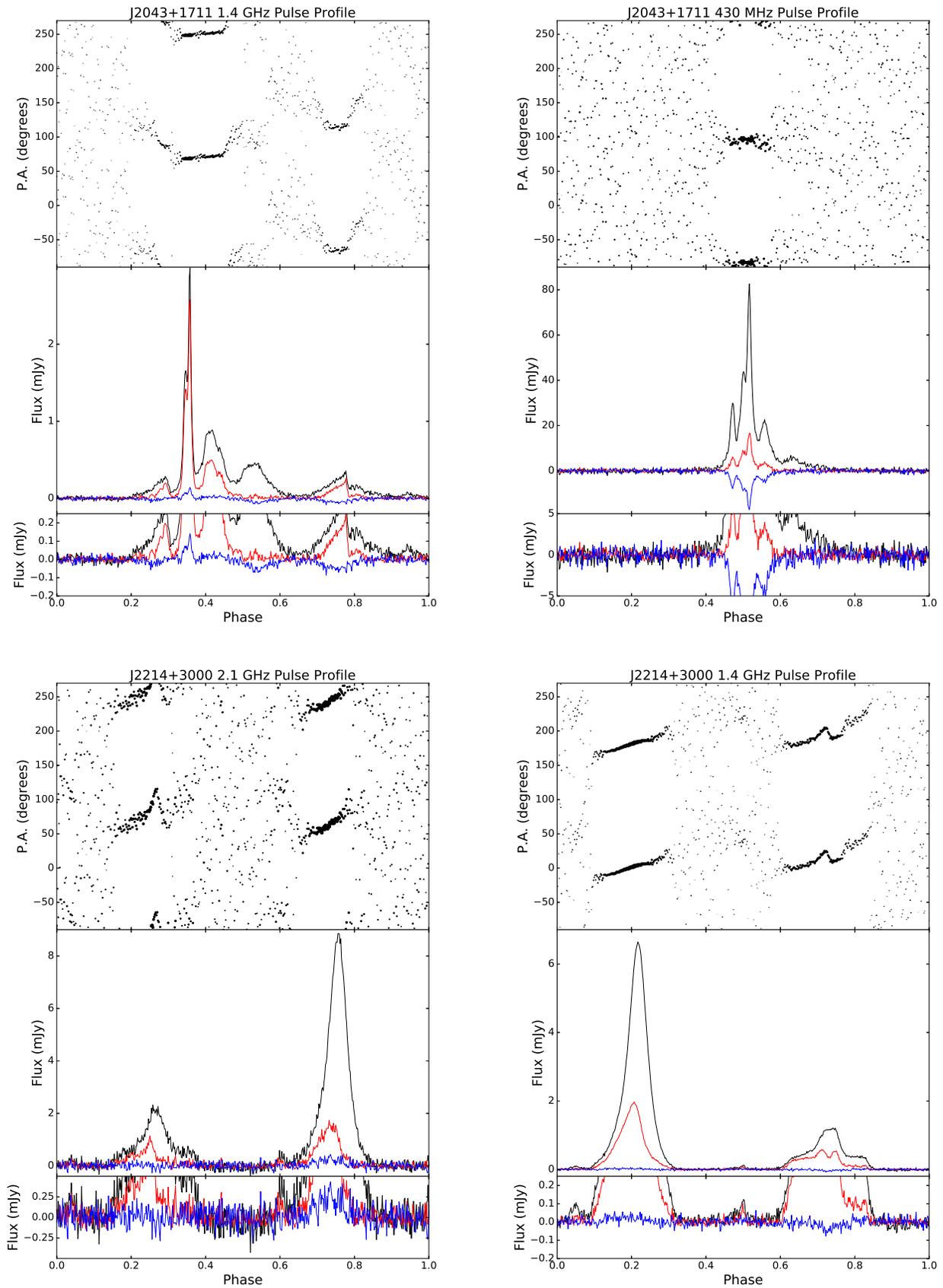

**Figure 14**: Same as Figure 1, for PSR J2043+1711 at 1.4 GHz and 430 MHz, and PSR J2214+3000 at 2.1 and 1.4 GHz.

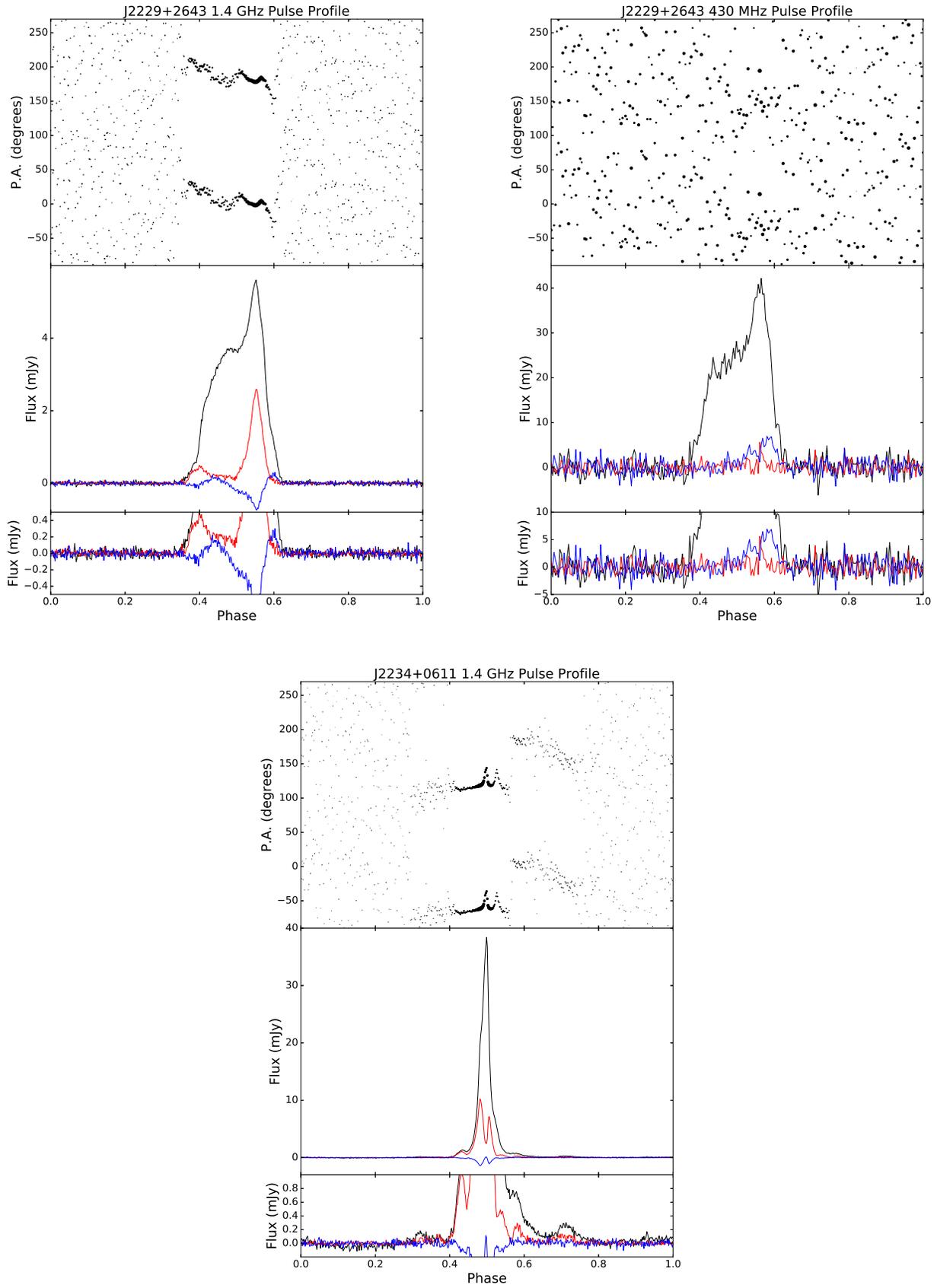

**Figure 15**: Same as Figure 1, for PSR J2229+2643 at 1.4 GHz and 430 MHz, and PSR J2234+0611 at 1.4 GHz.

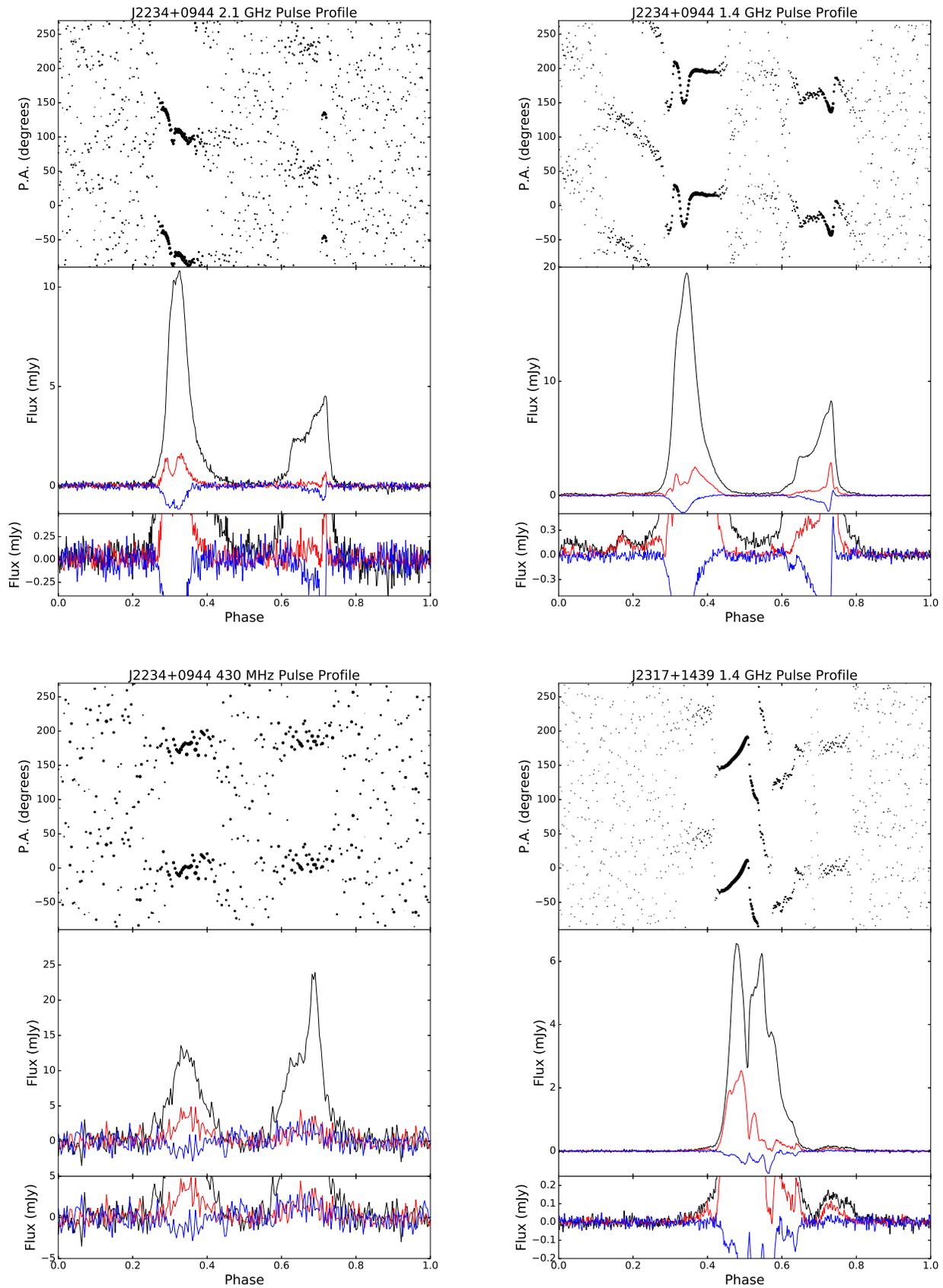

**Figure 16**: Same as Figure 1, for PSR J2234+0944 at 2.1 GHz, 1.4 GHz, and 430 MHz, and PSR J2317+1439 at 1.4 GHz.

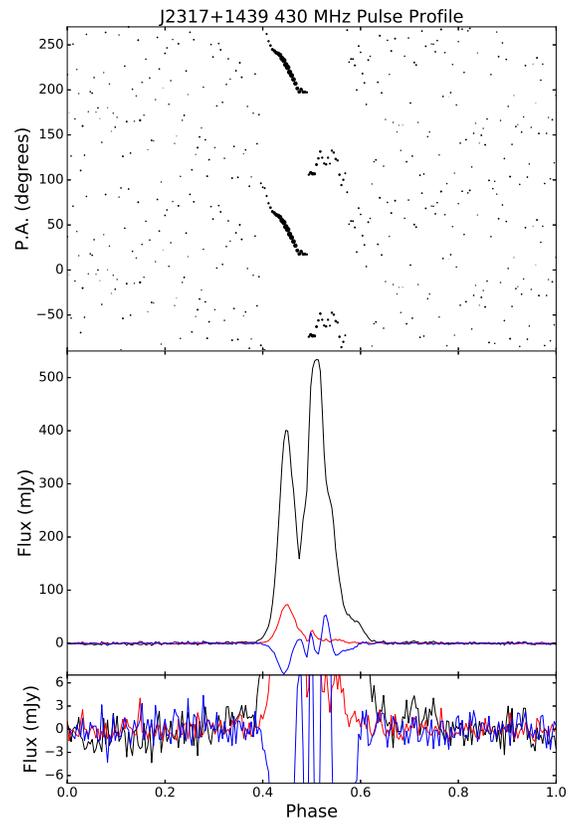

**Figure 17**: Same as Figure 1, for PSR J2317+1439 at 430 MHz.

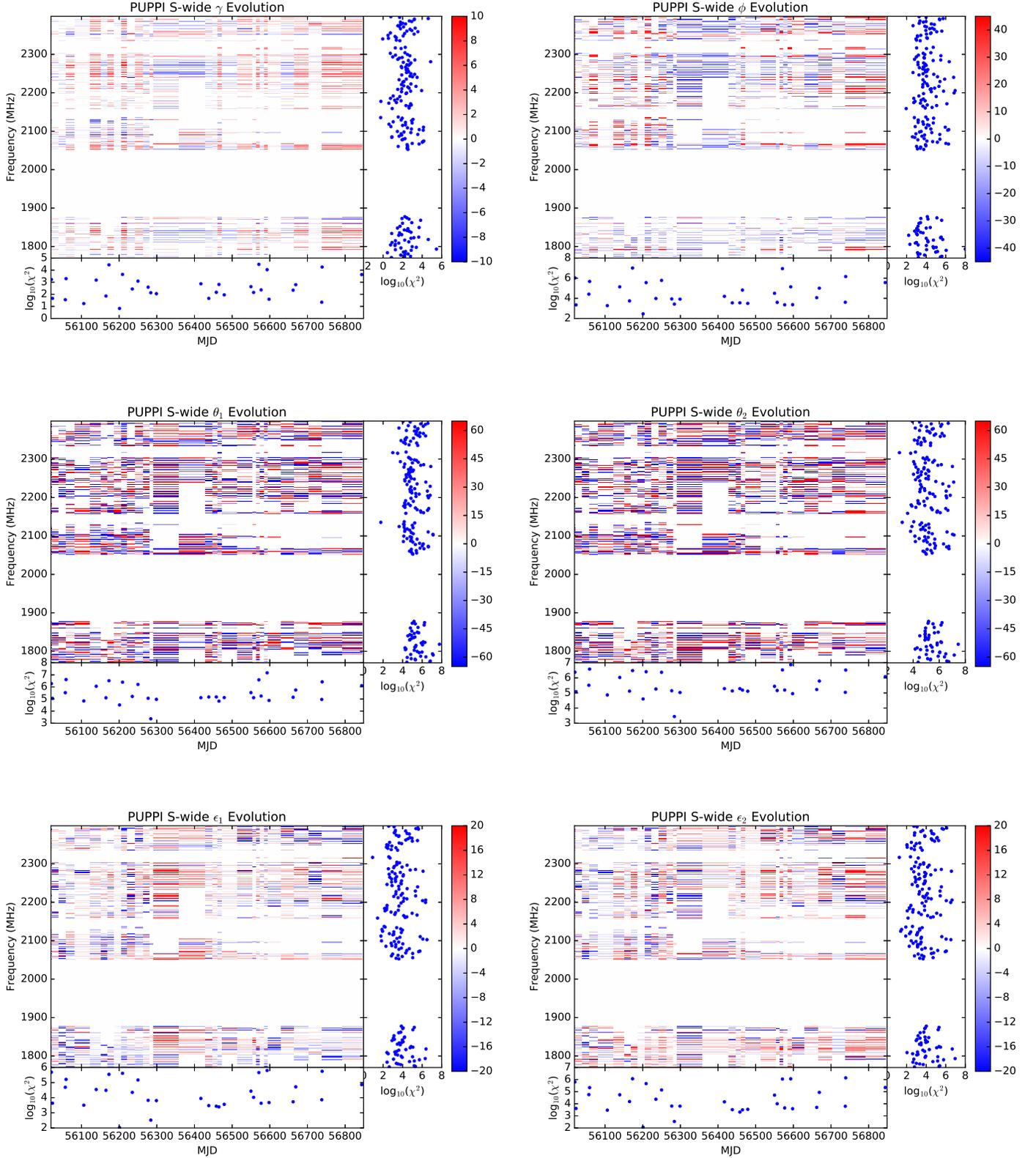

**Figure 18**: PRs of the S-wide receiver for both standard sources (see Equation 6 for a description of the parameters in the PR). For each color plot, the x-axis is MJD and the y-axis is observing frequency. To further highlight the differences, the color scales have been restricted. The lowest subplot of each plot shows the logarithm of the reduced $\chi^2$ of the given receiver parameter holding MJD constant, while the right-post subplot of each plot shows the logarithm of the reduced $\chi^2$ of the given receiver parameter in a specific frequency channel. Recall that the PRs are derived from observations that have already been calibrated with the MEM solution, and therefore the above plots do not show the absolute numerical values for each parameter versus frequency and MJD, but rather the difference between each parameter and its nominal MEM value over frequency and MJD. All angles are expressed in degrees.

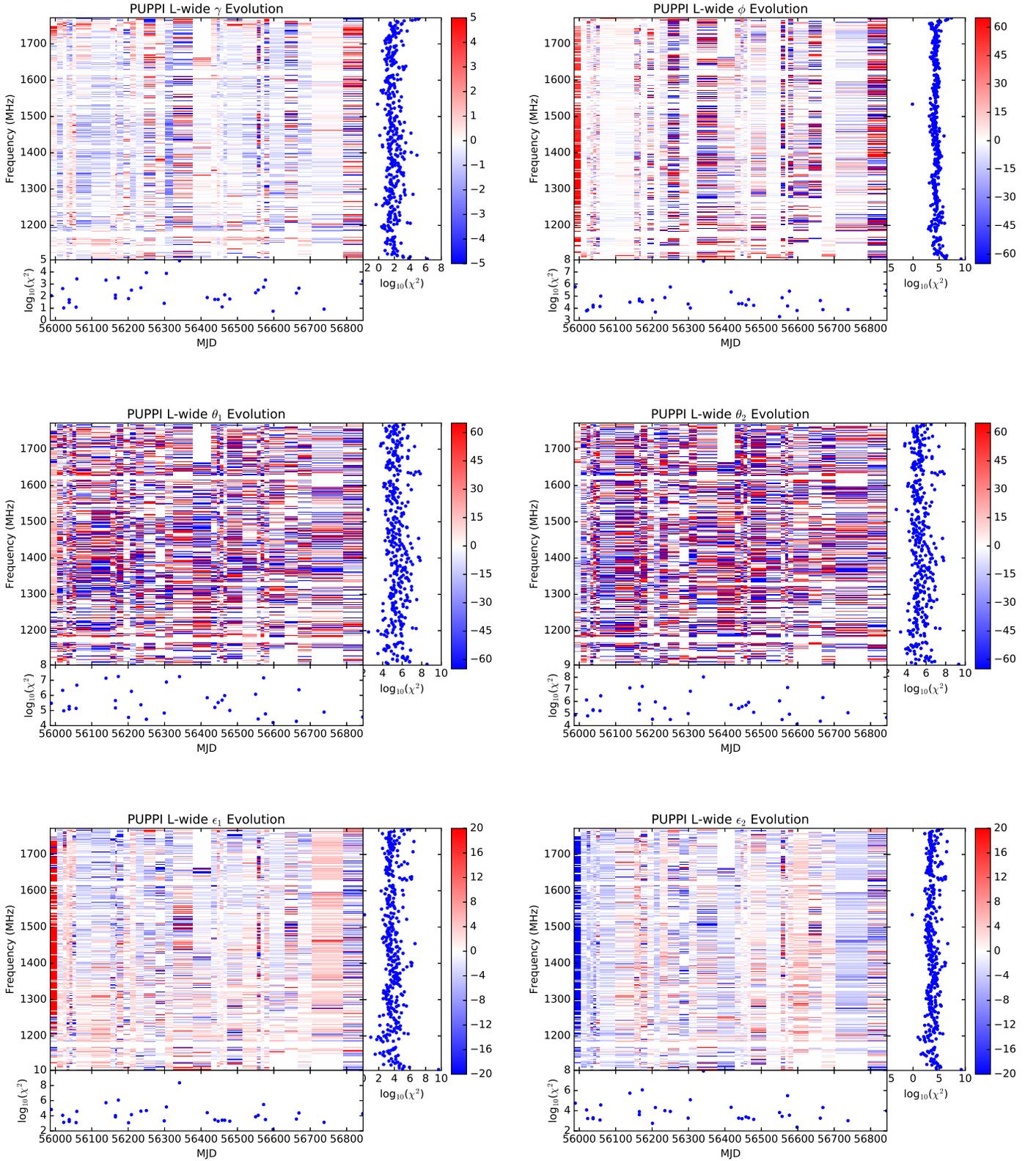

**Figure 19**: PRs of the L-wide receiver for both standard sources (see Equation 6 for a description of the parameters in the PR). For each color plot, the x-axis is MJD and the y-axis is observing frequency. To further highlight the differences, the color scales have been restricted. The lowest subplot of each plot shows the logarithm of the reduced $\chi^2$ of the given receiver parameter holding MJD constant, while the right-post subplot of each plot shows the logarithm of the reduced $\chi^2$ of the given receiver parameter in a specific frequency channel. Recall that the PRs are derived from observations that have already been calibrated with the MEM solution, and therefore the above plots do not show the absolute numerical values for each parameter versus frequency and MJD, but rather the difference between each parameter and its nominal MEM value over frequency and MJD. All angles are expressed in degrees.

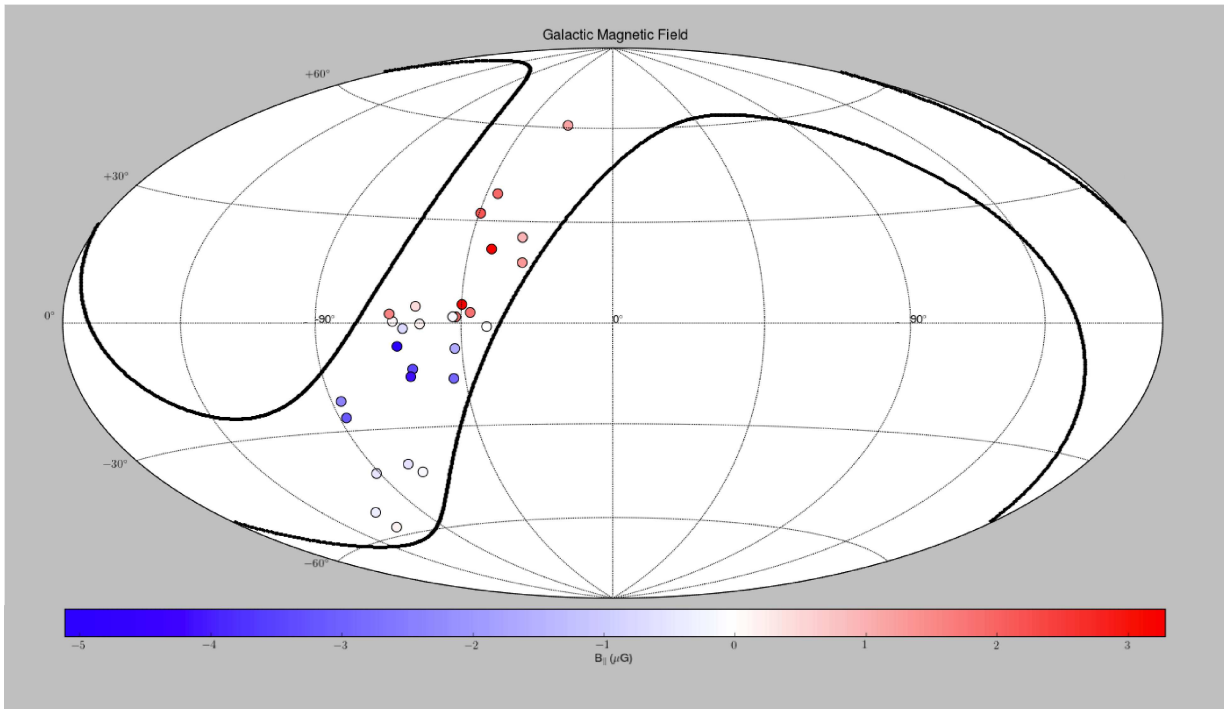

**Figure 20**: RM-derived values for the Galactic magnetic field parallel to the line of sight for each source as it appears on the sky in Galactic coordinates. Positive values denote a magnetic field pointing towards the Earth whereas negative values denote a magnetic field pointing away from the Earth. The asymmetry about a Galactic latitude of 0° agrees with Galactic magnetic field models.